\documentstyle[sprocl]{article}
\input epsf

\global\arraycolsep=2pt
\def\spose#1{\hbox to 0pt{#1\hss}}
\def\lsim{\mathrel{\spose{\lower 3pt\hbox{$\mathchar"218$}}
 \raise 2.0pt\hbox{$\mathchar"13C$}}}
\def\gsim{\mathrel{\spose{\lower 3pt\hbox{$\mathchar"218$}}
 \raise 2.0pt\hbox{$\mathchar"13E$}}}

\begin{document}

%%% start CERN preprint title page %%%%%%%%%%%%%
\begin{titlepage}

\begin{flushright}
CERN-TH/96-292\\
hep-ph/9610nnn
\end{flushright}

\vspace{2cm}

\begin{center}
\Large\bf Heavy-Quark Effective Theory
\end{center}

\vspace{2cm}

\begin{center}
Matthias Neubert\\
{\sl Theory Division, CERN, CH-1211 Geneva 23, Switzerland}
\end{center}

\vspace{1cm}

\begin{abstract}
We give an introduction to the heavy-quark effective theory and the
$1/m_Q$ expansion, which provide the modern framework for a
systematic, model-independent description of the properties and
decays of hadrons containing a heavy quark. We discuss the
applications of these concepts to spectroscopy and to the weak
decays of $B$ mesons.
\end{abstract}

\vspace{1.5cm}

\begin{center}
Invited talk presented at the\\
20th Johns Hopkins Workshop on Current Problems in Particle Theory\\
Heidelberg, Germany, 27-29 June 1996
\end{center}

\vspace{2cm}

\noindent
CERN-TH/96-292\\
October 1996
\vfil

\end{titlepage}

\thispagestyle{empty}
\vbox{}
\newpage

\setcounter{page}{1}

%%% end CERN preprint title page %%%%%%%%%%%%%

\title{HEAVY-QUARK EFFECTIVE THEORY}

\author{MATTHIAS NEUBERT}

\address{Theory Division, CERN, CH-1211 Geneva 23, Switzerland}

\maketitle\abstracts{
We give an introduction to the heavy-quark effective theory and the
$1/m_Q$ expansion, which provide the modern framework for a
systematic, model-independent description of the properties and
decays of hadrons containing a heavy quark. We discuss the
applications of these concepts to spectroscopy and to the weak
decays of $B$ mesons.}

\section{Introduction}
 
The weak decays of hadrons containing a heavy quark are employed for
tests of the Standard Model and measurements of its parameters. They
offer the most direct way to determine the weak mixing angles, to
test the unitarity of the Cabibbo--Kobayashi--Maskawa (CKM) matrix,
and to explore the physics of CP violation. At the same time,
hadronic weak decays also serve as a probe of that part of
strong-interaction phenomenology which is least understood: the
confinement of quarks and gluons inside hadrons.

\begin{figure}[htb]
   \epsfxsize=5cm
   \centerline{\epsffile{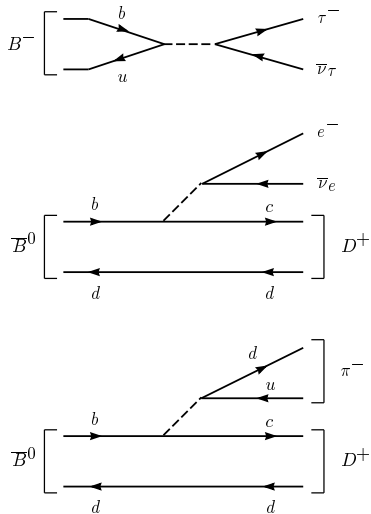}}
\caption{\label{fig:classes}
Examples of leptonic ($B^-\to\tau^-\bar\nu_\tau$), semileptonic
($\bar B^0\to D^+ e^-\bar\nu_e$), and non-leptonic ($\bar B^0\to
D^+\pi^-$) decays of $B$ mesons.}
\end{figure}

The structure of weak interactions in the Standard Model is rather
simple. Flavour-changing decays are mediated by the coupling of the
charged current to the $W$-boson field. At low energies, the
charged-current interaction gives rise to local four-fermion
couplings, whose strength is governed by the Fermi constant
\begin{equation}
   G_F = {g^2\over 4\sqrt{2} M_W^2} = 1.16639(2)~\mbox{GeV}^{-2} \,.
\end{equation}
According to the structure of the these interactions, the weak decays
of hadrons can be divided into three classes: leptonic decays, in
which the quarks of the decaying hadron annihilate each other and
only leptons appear in the final state; semileptonic decays, in which
both leptons and hadrons appear in the final state; and non-leptonic
decays, in which the final state consists of hadrons only.
Representative examples of these three types of decays are shown in
Fig.~\ref{fig:classes}.

\begin{figure}[htb]
   \epsfxsize=8.5cm
   \centerline{\epsffile{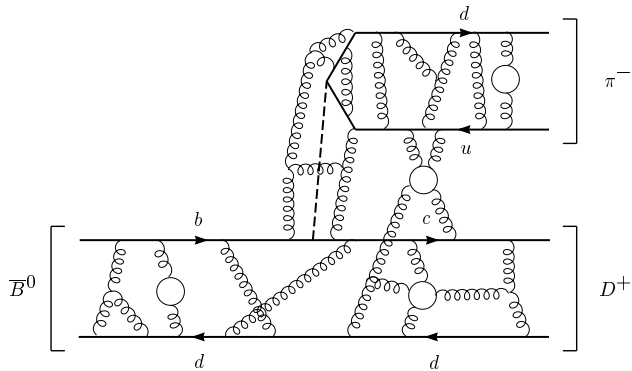}}
\caption{\label{fig:nonlep}
More realistic representation of a non-leptonic decay.}
\end{figure}

The simple quark-line graphs shown in this figure are a gross
oversimplification, however. In the real world, quarks are confined
inside hadrons, bound by the exchange of soft gluons. The simplicity
of the weak interactions is overshadowed by the complexity of the
strong interactions. A complicated interplay between the weak and
strong forces characterizes the phenomenology of hadronic weak
decays. As an example, a more realistic picture of a non-leptonic
decay is shown in Fig.~\ref{fig:nonlep}. Clearly, the complexity of
strong-interaction effects increases with the number of quarks
appearing in the final state. Bound-state effects in leptonic decays
can be lumped into a single parameter (a ``decay constant''), while
those in semileptonic decays are described by invariant form factors,
depending on the momentum transfer $q^2$ between the hadrons.
Approximate symmetries of the strong interactions help to constrain
the properties of these form factors. For non-leptonic decays, on the
other hand, we are still far from having a quantitative understanding
of strong-interaction effects even in the simplest decay modes.

Over the last decade, a lot of information on heavy-quark decays has
been collected in experiments at $e^+ e^-$ and hadron colliders. This
has led to a rather detailed knowledge of the flavour sector of the
Standard Model and many of the parameters associated with it. There
have been several great discoveries in this field, such as
$B^0$--$\bar B^0$ mixing~\cite{BBbar1,BBbar2}, charmless $B$
decays~\cite{btou1}$^-$\cite{Bpirho}, and rare decays induced by
penguin operators~\cite{BKstar,btos}. The experimental progress in
heavy-flavour physics has been accompanied by a significant progress
in theory, which was related to the discovery of heavy-quark symmetry
and the development of the heavy-quark effective theory (HQET). The
excitement about these developments is caused by the fact that they
allow (some) model-independent predictions in an area in which
``progress'' in theory often meant nothing more than the construction
of a new model, which could be used to estimate some
strong-interaction hadronic matrix elements. Here we explain the
physical picture behind heavy-quark symmetry and discuss the
construction, as well as simple applications, of the heavy-quark
expansion. Because of lack of time, we will have to focus on some
particularly important aspects. A more complete discussion of the
applications of this formalism to heavy-flavour phenomenology can be
found in some recent review articles~\cite{review}$^-$\cite{AFrev}.
The reader is also encouraged to consult the earlier review
papers~\cite{GeRev}$^-$\cite{Grorev} on the subject.

Hadronic bound states of a heavy quark with light constituents
(quarks, antiquarks and gluons) are characterized by a large
separation of mass scales: the heavy-quark mass $m_Q$ is much larger
than the mass scale $\Lambda_{\rm QCD}$ associated with the light
degrees of freedom. Equivalently, the Compton wave length of the
heavy quark ($\lambda_Q\sim 1/m_Q$) is much smaller than the size of
the hadron containing the heavy quark ($R_{\rm had}\sim
1/\Lambda_{\rm QCD}$). Our goal will be to separate the physics
associated with these two scales, in such a way that all dependence
on the heavy-quark mass becomes explicit. The framework in which to
perform this separation is the operator product expansion
(OPE)~\cite{Wils,Zimm}. The HQET provides us with a convenient
technical tool to construct the OPE. Before we start to explore the
details of this effective theory, however, we should mention two
important reasons why it is desirable to separate short- and
long-distance physics in the first place:
\begin{itemize}
\item
A technical reason is that after the separation of short- and
long-distance phenomena we can actually calculate a big portion of
the relevant physics (i.e.\ all short-distance effects) using
perturbation theory and renorma\-li\-za\-tion-group techniques. In
particular, in this way we will be able to control all logarithmic
dependence on the heavy-quark mass.
\item
An important physical reason is that, after the short-distance
physics has been separated, it may happen that the long-distance
physics simplifies due to the realization of approximate symmetries,
which imply non-trivial relations between observables.
\end{itemize}
The second point is particularly exciting, since it allows us to make
statements beyond the range of applicability of perturbation theory.
Notice that here we are not talking about symmetries of the full QCD
Lagrangian, such as its local gauge symmetry, but approximate
symmetries realized in a particular kinematic situation. In
particular, we will find that an approximate spin--flavour symmetry
is realized in systems in which a single heavy quark interacts with
light degrees of freedom by the exchange of soft gluons. 

At this point it is instructive to recall a more familiar example of
how approximate symmetries relate the long-distance physics of
several observables. The strong interactions of pions are severely
constrained by the approximate chiral symmetry of QCD. In a certain
kinematic regime, where the momenta of the pions are much less than
1~GeV (the scale of chiral-symmetry breaking), the long-distance
physics of scattering amplitudes is encoded in a few ``reduced matrix
elements'', such as the pion decay constant. An effective low-energy
theory called chiral perturbation theory provides a systematic
expansion of scattering amplitudes in powers of the pion momenta, and
thus helps to derive the relations between different scattering
amplitudes imposed by chiral symmetry~\cite{Leut}. We will find that
a similar situation holds for the case of heavy quarks. Heavy-quark
symmetry implies that, in the limit where $m_Q\gg\Lambda_{\rm QCD}$,
the long-distance physics of several observables is encoded in few
hadronic parameters, which can be defined in terms of operator matrix
elements in the HQET.

\section{Heavy-Quark Symmetry}

\subsection{The Physical Picture}
 
There are several reasons why the strong interactions of systems
containing heavy quarks are easier to understand than those of
systems containing only light quarks. The first is asymptotic
freedom, the fact that the effective coupling constant of QCD becomes
weak in processes with large momentum transfer, corresponding to
interactions at short-distance scales~\cite{Gros,Poli}. At large
distances, on the other hand, the coupling becomes
strong~\cite{Maria}, leading to non-perturbative phenomena such as
the confinement of quarks and gluons on a length scale $R_{\rm
had}\sim 1/\Lambda_{\rm QCD}\sim 1$~fm, which determines the size of
hadrons. Roughly speaking, $\Lambda_{\rm QCD}\sim 0.2$ GeV is the
energy scale that separates the regions of large and small coupling
constant. When the mass of a quark $Q$ is much larger than this
scale, it is called a heavy quark. The quarks of the Standard Model
fall naturally into two classes: up, down and strange are light
quarks, whereas charm, bottom and top are heavy
quarks.\footnote{Ironically, the top quark is of no relevance to our
discussion here, since it is too heavy to form hadronic bound states
before it decays.} For heavy quarks, the effective coupling constant
$\alpha_s(m_Q)$ is small, implying that on length scales comparable
to the Compton wavelength $\lambda_Q\sim 1/m_Q$ the strong
interactions are perturbative and similar to the electromagnetic
interactions. In fact, the quarkonium systems $(\bar QQ)$, whose size
is of order $\lambda_Q/\alpha_s(m_Q)\ll R_{\rm had}$, are very much
hydrogen-like.

Systems composed of a heavy quark and light constituents are
more complicated, however. The size of such systems is determined by
$R_{\rm had}$, and the typical momenta exchanged between the heavy
and light constituents are of order $\Lambda_{\rm QCD}$. The heavy
quark is surrounded by a most complicated, strongly interacting cloud
of light quarks, antiquarks, and gluons. In this case it is the fact
that $\lambda_Q\ll R_{\rm had}$, i.e.\ that the Compton wavelength of
the heavy quark is much smaller than the size of the hadron, which
leads to simplifications. To resolve the quantum numbers of the heavy
quark would require a hard probe; the soft gluons exchanged between
the heavy quark and the light constituents can only resolve distances
much larger than $\lambda_Q$. Therefore, the light degrees of freedom
are blind to the flavour (mass) and spin orientation of the heavy
quark. They experience only its colour field, which extends over
large distances because of confinement. In the rest frame of the
heavy quark, it is in fact only the electric colour field that is
important; relativistic effects such as colour magnetism vanish as
$m_Q\to\infty$. Since the heavy-quark spin participates in
interactions only through such relativistic effects, it decouples.
That the heavy-quark mass becomes irrelevant can be seen as follows:
As $m_Q\to\infty$, the heavy quark and the hadron that contains it
have the same velocity. In the rest frame of the hadron, the heavy
quark is at rest, too. The wave function of the light constituents
follows from a solution of the field equations of QCD subject to the
boundary condition of a static triplet source of colour at the
location of the heavy quark. This boundary condition is independent
of $m_Q$, and so is the solution for the configuration of the light
constituents.

It follows that, in the limit $m_Q\to\infty$, hadronic systems which
differ only in the flavour or spin quantum numbers of the heavy quark
have the same configuration of their light degrees of
freedom~\cite{Shu1}$^-$\cite{Isgu}. Although this observation still
does not allow us to calculate what this configuration is, it
provides relations between the properties of such particles as the
heavy mesons $B$, $D$, $B^*$ and $D^*$, or the heavy baryons
$\Lambda_b$ and $\Lambda_c$ (to the extent that corrections to the
infinite quark-mass limit are small in these systems). These
relations result from some approximate symmetries of the effective
strong interactions of heavy quarks at low energies. The
configuration of light degrees of freedom in a hadron containing a
single heavy quark with velocity $v$ does not change if this quark is
replaced by another heavy quark with different flavour or spin, but
with the same velocity. Both heavy quarks lead to the same static
colour field. For $N_h$ heavy-quark flavours, there is thus an SU$(2
N_h)$ spin--flavour symmetry group, under which the effective strong
interactions are invariant. These symmetries are in close
correspondence to familiar properties of atoms: The flavour symmetry
is analogous to the fact that different isotopes have the same
chemistry, since to a good approximation the wave function of the
electrons is independent of the mass of the nucleus. The electrons
only see the total nuclear charge. The spin symmetry is analogous to
the fact that the hyperfine levels in atoms are nearly degenerate.
The nuclear spin decouples in the limit $m_e/m_N\to 0$.

Heavy-quark symmetry is an approximate symmetry, and corrections
arise since the quark masses are not infinitely heavy. In many
respects, it is complementary to chiral symmetry, which arises in the
opposite limit of small quark masses. However, whereas chiral
symmetry is a symmetry of the QCD Lagrangian in the limit of
vanishing quark masses, heavy-quark symmetry is not a symmetry of the
Lagrangian (not even an approximate one), but rather a symmetry of an
effective theory, which is a good approximation of QCD in a certain
kinematic region. It is realized only in systems in which a heavy
quark interacts predominantly by the exchange of soft gluons. In such
systems the heavy quark is almost on shell; its momentum fluctuates
around the mass shell by an amount of order $\Lambda_{\rm QCD}$. The
corresponding fluctuations in the velocity of the heavy quark vanish
as $\Lambda_{\rm QCD}/m_Q\to 0$. The velocity becomes a conserved
quantity and is no longer a dynamical degree of freedom~\cite{Geor}.
Nevertheless, results derived on the basis of heavy-quark symmetry
are model-independent consequences of QCD in a well-defined limit.
The symmetry-breaking corrections can, at least in principle, be
studied in a systematic way. A convenient framework for analysing
these corrections is provided by the HQET. Before presenting a
detailed discussion of the formalism of this effective theory, we
shall first point out some of the important implications of
heavy-quark symmetry for the spectroscopy and weak decays of heavy
hadrons.

\subsection{Spectroscopic Implications}

The spin--flavour symmetry leads to many interesting relations
between the properties of hadrons containing a heavy quark. The most
direct consequences concern the spectroscopy of such
states~\cite{IsWi}. In the limit $m_Q\to\infty$, the spin of the
heavy quark and the total angular momentum $j$ of the light degrees
of freedom inside a hadron are separately conserved by the strong
interactions. Because of heavy-quark symmetry, the dynamics is
independent of the spin and mass of the heavy quark. Hadronic states
can thus be classified by the quantum numbers (flavour, spin, parity,
etc.) of the light degrees of freedom~\cite{AFal}. The spin symmetry
predicts that, for fixed $j\neq 0$, there is a doublet of degenerate
states with total spin $J=j\pm\frac{1}{2}$. The flavour symmetry
relates the properties of states with different heavy-quark flavour.

In general, the mass of a hadron $H_Q$ containing a heavy quark $Q$
obeys an expansion of the form
\begin{equation}\label{massexp}
   m_H = m_Q + \bar\Lambda + {\Delta m^2\over 2 m_Q}
   + O(1/m_Q^2) \,.
\end{equation}
The parameter $\bar\Lambda$ represents contributions arising from all
terms in the Lagrangian that are independent of the heavy-quark
mass~\cite{FNL}, whereas the quantity $\Delta m^2$ originates from
the terms of order $1/m_Q$ in the effective Lagrangian of the HQET.
For the moment, the detailed structure of these terms is of no
relevance; it will be discussed at length later on. For the
ground-state pseudoscalar and vector mesons, one can parametrize the
contributions from the $1/m_Q$ corrections in terms of two
quantities, $\lambda_1$ and $\lambda_2$, in such a way
that~\cite{FaNe}
\begin{equation}\label{FNrela}
   \Delta m^2 = -\lambda_1 + 2 \Big[ J(J+1) - \textstyle{3\over 2}
   \Big]\,\lambda_2 \,.
\end{equation}
Here $J$ is the total spin of the meson. The first term,
$-\lambda_1/2 m_Q$, arises from the kinetic energy of the heavy quark
inside the meson; the second term describes the interaction of the
heavy-quark spin with the gluon field. The hadronic parameters
$\bar\Lambda$, $\lambda_1$ and $\lambda_2$ are independent of $m_Q$.
They characterize the properties of the light constituents.

Consider, as a first example, the SU(3) mass splittings for heavy
mesons. The heavy-quark expansion predicts that
\begin{eqnarray}
   m_{B_S} - m_{B_d} &=& \bar\Lambda_s - \bar\Lambda_d
    + O(1/m_b) \,, \nonumber\\
   m_{D_S} - m_{D_d} &=& \bar\Lambda_s - \bar\Lambda_d
    + O(1/m_c) \,,
\end{eqnarray}
where we have indicated that the value of the parameter $\bar\Lambda$
depends on the flavour of the light quark. Thus, to the extent that
the charm and bottom quarks can both be considered sufficiently
heavy, the mass splittings should be similar in the two systems. This
prediction is confirmed experimentally, since~\cite{Joe}
\begin{eqnarray}
   m_{B_S} - m_{B_d} &=& (90\pm 3)~\mbox{MeV} \,, \nonumber\\
   m_{D_S} - m_{D_d} &=& (99\pm 1)~\mbox{MeV} \,.
\end{eqnarray}

As a second example, consider the spin splittings between the
ground-state pseudoscalar ($J=0$) and vector ($J=1$) mesons, which
are the members of the spin-doublet with $j=\frac{1}{2}$. The theory
predicts that
\begin{eqnarray}
   m_{B^*}^2 - m_B^2 &=& 4\lambda_2 + O(1/m_b) \,, \nonumber\\
   m_{D^*}^2 - m_D^2 &=& 4\lambda_2 + O(1/m_c) \,.
\end{eqnarray}
The data are compatible with this:
\begin{eqnarray}\label{VPexp}
   m_{B^*}^2 - m_B^2 &\simeq& 0.49~{\rm GeV}^2 \,, \nonumber\\
   m_{D^*}^2 - m_D^2 &\simeq& 0.55~{\rm GeV}^2 \,.
\end{eqnarray}
Assuming that the $B$ system is close to the heavy-quark limit, we
obtain the value
\begin{equation}\label{lam2val}
   \lambda_2\simeq 0.12~\mbox{GeV}^2
\end{equation}
for one of the hadronic parameters in (\ref{FNrela}). As we shall see
later, this quantity plays an important role in the phenomenology of
inclusive decays of heavy hadrons.

A third example is provided by the mass splittings between the
ground-state mesons and baryons containing a heavy quark. The HQET
predicts that
\begin{eqnarray}\label{barmes}
   m_{\Lambda_b} - m_B &=& \bar\Lambda_{\rm baryon}
    - \bar\Lambda_{\rm meson} + O(1/m_b) \,, \nonumber\\
   m_{\Lambda_c} - m_D &=& \bar\Lambda_{\rm baryon}
    - \bar\Lambda_{\rm meson} + O(1/m_c) \,.
\end{eqnarray}
This is again consistent with the experimental results
\begin{eqnarray}
   m_{\Lambda_b} - m_B &=& (346\pm 6)~\mbox{MeV} \,, \nonumber\\
   m_{\Lambda_c} - m_D &=& (416\pm 1)~\mbox{MeV} \,,
\end{eqnarray}
although in this case the data indicate sizeable symmetry-breaking
corrections. For the mass of the $\Lambda_b$ baryon, we have used the
value
\begin{equation}\label{Lbmass}
   m_{\Lambda_b} = (5625\pm 6)~\mbox{MeV} \,,
\end{equation}
which is obtained by averaging the result~\cite{Joe} $m_{\Lambda_b}=
(5639\pm 15)$~MeV with the value $m_{\Lambda_b}=(5623\pm 5\pm 4)$~MeV
reported by the CDF Collaboration~\cite{CDFmass}. The dominant
correction to the relations (\ref{barmes}) comes from the
contribution of the chromo-magnetic interaction to the masses of the
heavy mesons,\footnote{Because of the spin symmetry, there is no such
contribution to the masses of the $\Lambda_Q$ baryons.} which adds a
term $3\lambda_2/2 m_Q$ on the right-hand side. Including this term,
we obtain the refined prediction that the values of the following two
quantities should be close to each other:
\begin{eqnarray}
   m_{\Lambda_b} - m_B - {3\lambda_2\over 2 m_B}
   &=& (312\pm 6)~\mbox{MeV} \,, \nonumber\\
   m_{\Lambda_c} - m_D - {3\lambda_2\over 2 m_D}
   &=& (320\pm 1)~\mbox{MeV}
\end{eqnarray}
This is clearly satisfied by the data.

The mass formula (\ref{massexp}) can also be used to derive
information on the heavy-quark (pole) masses from the observed hadron
masses. Introducing the ``spin-averaged'' meson masses
$\overline{m}_B=\frac{1}{4}\,(m_B+3 m_{B^*})\simeq 5.31$~GeV and
$\overline{m}_D=\frac{1}{4}\,(m_D+3 m_{D^*})\simeq 1.97$~GeV, we find
that
\begin{equation}\label{mbmc}
   m_b-m_c = (\overline{m}_B-\overline{m}_D)\,\bigg\{
   1 - {\lambda_1\over 2\overline{m}_B\overline{m}_D}
   + O(1/m_Q^3) \bigg\} \,,
\end{equation}
where $O(1/m_Q^3)$ is used as a generic notation representing terms
suppressed by three powers of the $b$- or $c$-quark masses. Using
theoretical estimates for the parameter $\lambda_1$, which lie in the
range~\cite{lam1}$^-$\cite{virial}
\begin{equation}\label{lam1}
   \lambda_1 = -(0.3\pm 0.2)~\mbox{GeV}^2 \,,
\end{equation}
this relation leads to
\begin{equation}\label{mbmcval}
   m_b - m_c = (3.39\pm 0.03\pm 0.03)~\mbox{GeV} \,,
\end{equation}
where the first error reflects the uncertainty in the value of
$\lambda_1$, and the second one takes into account unknown
higher-order corrections.

\subsection{Exclusive Semileptonic Decays}
\label{sec:3}

Semileptonic decays of $B$ mesons have received a lot of attention in
recent years. The decay channel $\bar B\to D^*\ell\,\bar\nu$ has the
largest branching fraction of all $B$-meson decay modes. From a
theoretical point of view, semileptonic decays are simple enough to
allow for a reliable, quantitative description. The analysis of these
decays provides much information about the strong forces that bind
the quarks and gluons into hadrons. Heavy-quark symmetry implies
relations between the weak decay form factors of heavy mesons, which
are of particular interest. These relations have been derived by
Isgur and Wise~\cite{Isgu}, generalizing ideas developed by Nussinov
and Wetzel~\cite{Nuss}, and by Voloshin and Shifman~\cite{Vol1,Vol2}.

\begin{figure}[htb]
   \epsfxsize=7cm
   \centerline{\epsffile{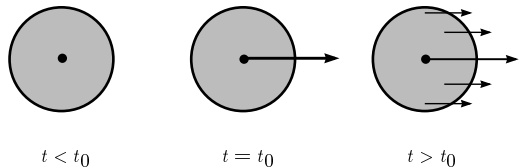}}
\caption{\label{fig:3.3}
Elastic transition induced by an external heavy-quark current.}
\end{figure}

Consider the elastic scattering of a $B$ meson, $\bar B(v)\to\bar
B(v')$, induced by a vector current coupled to the $b$ quark. Before
the action of the current, the light degrees of freedom inside the
$B$ meson orbit around the heavy quark, which acts as a static source
of colour. On average, the $b$ quark and the $B$ meson have the same
velocity $v$. The action of the current is to replace instantaneously
(at time $t=t_0$) the colour source by one moving at a velocity $v'$,
as indicated in Fig.~\ref{fig:3.3}. If $v=v'$, nothing happens; the
light degrees of freedom do not realize that there was a current
acting on the heavy quark. If the velocities are different, however,
the light constituents suddenly find themselves interacting with a
moving colour source. Soft gluons have to be exchanged to rearrange
them so as to form a $B$ meson moving at velocity $v'$. This
rearrangement leads to a form-factor suppression, which reflects the
fact that as the velocities become more and more different, the
probability for an elastic transition decreases. The important
observation is that, in the limit $m_b\to\infty$, the form factor can
only depend on the Lorentz boost $\gamma = v\cdot v'$ that connects
the rest frames of the initial- and final-state mesons. Thus, in this
limit a dimensionless probability function $\xi(v\cdot v')$ describes
the transition. It is called the Isgur--Wise function~\cite{Isgu}. In
the HQET, which provides the appropriate framework for taking the
limit $m_b\to\infty$, the hadronic matrix element describing the
scattering process can thus be written as
\begin{equation}\label{elast}
   {1\over m_B}\,\langle\bar B(v')|\,\bar b_{v'}\gamma^\mu b_v\,
   |\bar B(v)\rangle = \xi(v\cdot v')\,(v+v')^\mu \,.
\end{equation}
Here, $b_v$ and $b_{v'}$ are the velocity-dependent heavy-quark
fields of the HQET, whose precise definition will be discussed later.
It is important that the function $\xi(v\cdot v')$ does not depend on
$m_b$. The factor $1/m_B$ on the left-hand side compensates for a
trivial dependence on the heavy-meson mass caused by the relativistic
normalization of meson states, which is conventionally taken to be
\begin{equation}\label{nonrelnorm}
   \langle\bar B(p')|\bar B(p)\rangle = 2 m_B v^0\,(2\pi)^3\,
   \delta^3(\vec p-\vec p\,') \,.
\end{equation}
Note that there is no term proportional to $(v-v')^\mu$ in
(\ref{elast}). This can be seen by contracting the matrix element
with $(v-v')_\mu$, which must give zero since $\rlap/v\,b_v = b_v$ and
$\bar b_{v'}\rlap/v' = \bar b_{v'}$.

It is more conventional to write the above matrix element in terms of
an elastic form factor $F_{\rm el}(q^2)$ depending on the momentum
transfer $q^2=(p-p')^2$:
\begin{equation}
   \langle\bar B(v')|\,\bar b\,\gamma^\mu b\,|\bar B(v)\rangle
   = F_{\rm el}(q^2)\,(p+p')^\mu \,,
\end{equation}
where $p^(\phantom{}'\phantom{}^)=m_B v^(\phantom{}'\phantom{}^)$.
Comparing this with (\ref{elast}), we find that
\begin{equation}
   F_{\rm el}(q^2) = \xi(v\cdot v') \,, \qquad
   q^2 = -2 m_B^2 (v\cdot v'-1) \,.
\end{equation}
Because of current conservation, the elastic form factor is
normalized to unity at $q^2=0$. This condition implies the
normalization of the Isgur--Wise function at the kinematic point
$v\cdot v'=1$, i.e.\ for $v=v'$:
\begin{equation}\label{Jcons2}
   \xi(1) = 1 \,.
\end{equation}
It is in accordance with the intuitive argument that the probability
for an elastic transition is unity if there is no velocity change.
Since for $v=v'$ the daughter meson is at rest in the rest frame of
the parent meson, the point $v\cdot v'=1$ is referred to as the
zero-recoil limit.

We can now use the flavour symmetry to replace the $b$ quark in the
final-state meson by a $c$ quark, thereby turning the $B$ meson into
a $D$ meson. Then the scattering process turns into a weak decay
process. In the infinite-mass limit, the replacement $b_{v'}\to
c_{v'}$ is a symmetry transformation, under which the effective
Lagrangian is invariant. Hence, the matrix element
\begin{equation}
   {1\over\sqrt{m_B m_D}}\,\langle D(v')|\,\bar c_{v'}\gamma^\mu
   b_v\,|\bar B(v)\rangle = \xi(v\cdot v')\,(v+v')^\mu
\end{equation}
is still determined by the same function $\xi(v\cdot v')$. This is
interesting, since in general the matrix element of a
flavour-changing current between two pseudoscalar mesons is described
by two form factors:
\begin{equation}
   \langle D(v')|\,\bar c\,\gamma^\mu b\,|\bar B(v)\rangle
   = f_+(q^2)\,(p+p')^\mu - f_-(q^2)\,(p-p')^\mu \,.
\end{equation}
Comparing the above two equations, we find that
\begin{eqnarray}\label{inelast}
   f_\pm(q^2) &=& {m_B\pm m_D\over 2\sqrt{m_B m_D}}\,\xi(v\cdot v')
    \,, \nonumber\\
   q^2 &=& m_B^2 + m_D^2 - 2 m_B m_D\,v\cdot v' \,.
\end{eqnarray}
Thus, the heavy-quark flavour symmetry relates two a priori
independent form factors to one and the same function. Moreover, the
normalization of the Isgur--Wise function at $v\cdot v'=1$ now
implies a non-trivial normalization of the form factors $f_\pm(q^2)$
at the point of maximum momentum transfer, $q_{\rm max}^2=
(m_B-m_D)^2$:
\begin{equation}
   f_\pm(q_{\rm max}^2) = {m_B\pm m_D\over 2\sqrt{m_B m_D}} \,.
\end{equation}

The heavy-quark spin symmetry leads to additional relations among
weak decay form factors. It can be used to relate matrix elements
involving vector mesons to those involving pseudoscalar mesons. A
vector meson with longitudinal polarization is related to a
pseudoscalar meson by a rotation of the heavy-quark spin. Hence, the
spin-symmetry transformation $c_{v'}^\Uparrow\to c_{v'}^\Downarrow$
relates $\bar B\to D$ with $\bar B\to D^*$ transitions. The result of
this transformation is~\cite{Isgu}:
\begin{eqnarray}
   {1\over\sqrt{m_B m_{D^*}}}\,
   \langle D^*(v',\varepsilon)|\,\bar c_{v'}\gamma^\mu b_v\,
   |\bar B(v)\rangle &=& i\epsilon^{\mu\nu\alpha\beta}\,
    \varepsilon_\nu^*\,v'_\alpha v_\beta\,\,\xi(v\cdot v') \,,
    \nonumber\\
   {1\over\sqrt{m_B m_{D^*}}}\,
   \langle D^*(v',\varepsilon)|\,\bar c_{v'}\gamma^\mu\gamma_5\,
   b_v\,|\bar B(v)\rangle &=& \Big[ \varepsilon^{*\mu}\,(v\cdot v'+1)
    - v'^\mu\,\varepsilon^*\!\cdot v \Big]\,\xi(v\cdot v') \,,
    \nonumber\\
\end{eqnarray}
where $\varepsilon$ denotes the polarization vector of the $D^*$
meson. Once again, the matrix elements are completely described in
terms of the Isgur--Wise function. Now this is even more remarkable,
since in general four form factors, $V(q^2)$ for the vector current,
and $A_i(q^2)$, $i=0,1,2$, for the axial current, are required to
parametrize these matrix elements. In the heavy-quark limit, they
obey the relations~\cite{Neu1}
\begin{eqnarray}\label{PVff}
   {m_B+m_{D^*}\over 2\sqrt{m_B m_{D^*}}}\,\xi(v\cdot v')
   &=& V(q^2) = A_0(q^2) = A_1(q^2) \nonumber\\
   &=& \bigg[ 1 - {q^2\over(m_B+m_D)^2} \bigg]^{-1}\,A_1(q^2) \,,
    \nonumber\\
   \phantom{ \Bigg[ }
   q^2 &=& m_B^2 + m_{D^*}^2 - 2 m_B m_{D^*}\,v\cdot v' \,.
\end{eqnarray}

Equations (\ref{inelast}) and (\ref{PVff}) summarize the relations
imposed by heavy-quark symmetry on the weak decay form factors
describing the semileptonic decay processes $\bar B\to
D\,\ell\,\bar\nu$ and $\bar B\to D^*\ell\,\bar\nu$. These relations
are model-independent consequences of QCD in the limit where $m_b,
m_c\gg\Lambda_{\rm QCD}$. They play a crucial role in the
determination of the CKM matrix element $|V_{cb}|$. In terms of the
recoil variable $w=v\cdot v'$, the differential semileptonic decay
rates in the heavy-quark limit become~\cite{Vcb}:
\begin{eqnarray}\label{rates}
   {{\rm d}\Gamma(\bar B\to D\,\ell\,\bar\nu)\over{\rm d}w}
   &=& {G_F^2\over 48\pi^3}\,|V_{cb}|^2\,(m_B+m_D)^2\,m_D^3\,
    (w^2-1)^{3/2}\,\xi^2(w) \,, \nonumber\\
   \phantom{ \Bigg[ }
   {{\rm d}\Gamma(\bar B\to D^*\ell\,\bar\nu)\over{\rm d}w}
   &=& {G_F^2\over 48\pi^3}\,|V_{cb}|^2\,(m_B-m_{D^*})^2\,
    m_{D^*}^3\,\sqrt{w^2-1}\,(w+1)^2 \nonumber\\
   &&\times \Bigg[ 1 + {4w\over w+1}\,
    {m_B^2 - 2 w\,m_B m_{D^*} + m_{D^*}^2\over(m_B-m_{D^*})^2}
    \Bigg]\,\xi^2(w) \,.
\end{eqnarray}
These expressions receive symmetry-breaking corrections, since the
masses of the heavy quarks are not infinitely heavy. Perturbative
corrections of order $\alpha_s^n(m_Q)$ can be calculated order by
order in perturbation theory. A more difficult task is to control
the non-perturbative power corrections of order $(\Lambda_{\rm
QCD}/m_Q)^n$. The HQET provides a systematic framework for analysing
these corrections. For the case of weak-decay form factors, the
analysis of the $1/m_Q$ corrections was performed by
Luke~\cite{Luke}. Later, Falk and the present author have also
analysed the structure of $1/m_Q^2$ corrections for both meson and
baryon weak decay form factors~\cite{FaNe}. We shall not discuss
these rather technical issues in detail, but only mention the most
important result of Luke's analysis. It concerns the zero-recoil
limit, where an analogue of the Ademollo--Gatto theorem~\cite{AGTh}
can be proved. This is Luke's theorem~\cite{Luke}, which states that
the matrix elements describing the leading $1/m_Q$ corrections to
weak decay amplitudes vanish at zero recoil. This theorem is valid to
all orders in perturbation theory~\cite{FaNe,Neu7,ChGr}. Most
importantly, it protects the $\bar B\to D^*\ell\,\bar\nu$ decay rate
from receiving first-order $1/m_Q$ corrections at zero
recoil~\cite{Vcb}. (A similar statement is not true for the decay
$\bar B\to D\,\ell\,\bar\nu$, however. The reason is simple but
somewhat subtle. Luke's theorem protects only those form factors not
multiplied by kinematic factors that vanish for $v=v'$. By angular
momentum conservation, the two pseudoscalar mesons in the decay $\bar
B\to D\,\ell\,\bar\nu$ must be in a relative $p$ wave, and hence the
amplitude is proportional to the velocity $|\vec v_D|$ of the $D$
meson in the $B$-meson rest frame. This leads to a factor $(w^2-1)$
in the decay rate. In such a situation, form factors that are
kinematically suppressed can contribute~\cite{Neu1}.)

\subsection{Model-Independent Determination of $|V_{cb}|$}
 
We will now discuss the most
important application of the HQET in the context of semileptonic
decays of $B$ mesons. A model-independent determination of the CKM
matrix element $|V_{cb}|$ based on heavy-quark symmetry can be
obtained by measuring the recoil spectrum of $D^*$ mesons produced in
$\bar B\to D^*\ell\,\bar\nu$ decays~\cite{Vcb}. In the heavy-quark
limit, the differential decay rate for this process has been given in
(\ref{rates}). In order to allow for corrections to that limit, we
write
\begin{eqnarray}
   {{\rm d}\Gamma(\bar B\to D^*\ell\,\bar\nu)\over{\rm d}w}
   &=& {G_F^2\over 48\pi^3}\,(m_B-m_{D^*})^2\,m_{D^*}^3
    \sqrt{w^2-1}\,(w+1)^2 \nonumber\\
   &&\mbox{}\times \Bigg[ 1 + {4w\over w+1}\,
    {m_B^2-2w\,m_B m_{D^*} + m_{D^*}^2\over(m_B - m_{D^*})^2}
    \Bigg]\,|V_{cb}|^2\,{\cal{F}}^2(w) \,, \nonumber\\
\end{eqnarray}
where the hadronic form factor ${\cal F}(w)$ coincides with the
Isgur--Wise function up to symmetry-breaking corrections of order
$\alpha_s(m_Q)$ and $\Lambda_{\rm QCD}/m_Q$. The idea is to measure
the product $|V_{cb}|\,{\cal F}(w)$ as a function of $w$, and to
extract $|V_{cb}|$ from an extrapolation of the data to the
zero-recoil point $w=1$, where the $B$ and the $D^*$ mesons have a
common rest frame. At this kinematic point, heavy-quark symmetry
helps to calculate the normalization ${\cal F}(1)$ with small and
controlled theoretical errors. Since the range of $w$ values
accessible in this decay is rather small ($1<w<1.5$), the
extrapolation can be done using an expansion around $w=1$:
\begin{equation}\label{Fexp}
   {\cal F}(w) = {\cal F}(1)\,\Big[ 1 - \hat\varrho^2\,(w-1)
   + \dots \Big] \,.
\end{equation}
The slope $\hat\varrho^2$ is treated as a fit parameter.

\begin{figure}[htb]
   \epsfxsize=8cm
   \vspace{0.3cm}
   \centerline{\epsffile{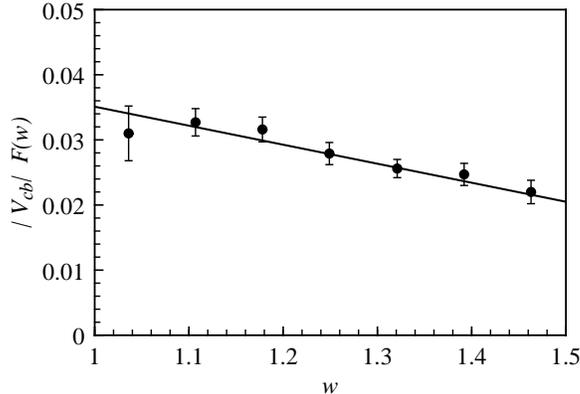}}
   \vspace{-0.3cm}
\caption{\label{fig:CLVcb}
CLEO data for the product $|V_{cb}|\,{\cal F}(w)$, as extracted from
the recoil spectrum in $\bar B\to D^*\ell\,\bar\nu$
decays~\protect\cite{CLEOVcb}. The line shows a linear fit to the
data.}
\end{figure}

Measurements of the recoil spectrum have been performed first by the
ARGUS~\cite{ARGVcb} and CLEO~\cite{CLEOVcb} Collaborations in
experiments operating at the $\Upsilon(4s)$ resonance, and more
recently by the ALEPH~\cite{ALEVcb} and DELPHI~\cite{DELVcb}
Collaborations at LEP. As an example, Fig.~\ref{fig:CLVcb} shows the
data reported by the CLEO Collaboration. The results obtained by the
various experimental groups from a linear fit to their data are
summarized in Table~\ref{tab:Vcb}. The weighted average of these
results is
\begin{eqnarray}\label{VcbFraw}
   |V_{cb}|\,{\cal F}(1) &=& (34.6\pm 1.7)\times 10^{-3} \,,
    \nonumber\\
   \hat\varrho^2 &=& 0.82\pm 0.09 \,.
\end{eqnarray}
The effect of a positive curvature of the form factor has been
investigated by Stone~\cite{Stone}, who finds that the value of
$|V_{cb}|\,{\cal F}(1)$ may change by up to $+4\%$. We thus increase
the above value by $(2\pm 2)\%$ and quote the final result as
\begin{equation}\label{VcbF}
   |V_{cb}|\,{\cal F}(1) = (35.3\pm 1.8)\times 10^{-3} \,.
\end{equation}
In future analyses, the extrapolation to zero recoil should be
performed including higher-order terms in the expansion (\ref{Fexp}).
It can be shown in a model-independent way that the shape of the form
factor is highly constrained by analyticity and unitarity
requirements~\cite{Boyd2,Capr}. In particular, the curvature at $w=1$
is strongly correlated with the slope of the form factor. For the
value of $\hat\varrho^2$ given in (\ref{VcbFraw}), one obtains a
small positive curvature~\cite{Capr}, in agreement with the
assumption made in Ref.~48.
%\citelow{Stone}

\begin{table}[htb]
\caption{\label{tab:Vcb}
Values for $|V_{cb}|\,{\cal F}(1)$ (in units of $10^{-3}$) and
$\hat\varrho^2$ extracted from measurements of the recoil spectrum in
$\bar B\to D^*\ell\,\bar\nu$ decays}
\vspace{0.4cm}
\begin{center}
\begin{tabular}{|l|c|c|}\hline
\rule[-0.15cm]{0cm}{0.65cm} & $|V_{cb}|\,{\cal F}(1)~(10^{-3})$ &
 $\hat\varrho^2$ \\
\hline
ARGUS  & $38.8\pm 4.3\pm 2.5$ & $1.17\pm 0.22\pm 0.06$ \\
CLEO   & $35.1\pm 1.9\pm 2.0$ & $0.84\pm 0.12\pm 0.08$ \\
ALEPH  & $31.4\pm 2.3\pm 2.5$ & $0.39\pm 0.21\pm 0.12$ \\
DELPHI & $35.0\pm 1.9\pm 2.3$ & $0.81\pm 0.16\pm 0.10$ \\
\hline
\end{tabular}
\end{center}
\end{table}

Heavy-quark symmetry implies that the general structure of the
symmetry-breaking corrections to the form factor at zero recoil
is~\cite{Vcb}
\begin{equation}
   {\cal F}(1) = \eta_A\,\bigg( 1 + 0 \times
   {\Lambda_{\rm QCD}\over m_Q}
   + \mbox{const} \times {\Lambda_{\rm QCD}^2\over m_Q^2}
   + \dots \bigg)
   \equiv \eta_A\,(1+\delta_{1/m^2}) \,,
\end{equation}
where $\eta_A$ is a short-distance correction arising from the
(finite) renormalization of the flavour-changing axial current at
zero recoil, and $\delta_{1/m^2}$ parametrizes second-order (and
higher) power corrections. The absence of first-order power
corrections at zero recoil is a consequence of Luke's
theorem~\cite{Luke}. The one-loop expression for $\eta_A$ has been
known for a long time~\cite{Pasc,Vol2,QCD1}:
\begin{equation}\label{etaA1}
   \eta_A = 1 + {\alpha_s(M)\over\pi}\,\bigg(
   {m_b+m_c\over m_b-m_c}\,\ln{m_b\over m_c} - {8\over 3} \bigg)
   \simeq 0.96 \,.
\end{equation}
The scale $M$ in the running coupling constant can be fixed by
adopting the prescription of Brodsky, Lepage and Mackenzie
(BLM)~\cite{BLM}, according to which it is identified with the
average virtuality of the gluon in the one-loop diagrams that
contribute to $\eta_A$. If $\alpha_s(M)$ is defined in the modified
minimal subtraction ($\overline{\mbox{\sc ms}}$) scheme, the result
is~\cite{etaVA} $M\simeq 0.51\sqrt{m_c m_b}$. Several estimates of
higher-order corrections to $\eta_A$ have been discussed. The
next-to-leading order resummation of logarithms of the type
$[\alpha_s\ln( m_b/m_c)]^n$ leads to~\cite{FaGr,QCD2} $\eta_A\simeq
0.985$. On the other hand, the resummation of ``renormalon-chain''
contributions of the form $\beta_0^{n-1}\alpha_s^n$, where $\beta_0$
is the first coefficient of the QCD $\beta$-function,
gives~\cite{flow,BaBB} $\eta_A\simeq 0.945$. Using these partial
resummations to estimate the uncertainty results in $\eta_A =
0.965\pm 0.020$. Recently, Czarnecki has improved this estimate by
calculating $\eta_A$ at two-loop order~\cite{Czar}. His result,
\begin{equation}
   \eta_A = 0.960\pm 0.007 \,,
\end{equation}
is in excellent agreement with the BLM-improved one-loop estimate
(\ref{etaA1}). Here the error is taken to be the size of the two-loop
correction.

The analysis of the power corrections $\delta_{1/m^2}$ is more
difficult, since it cannot rely on perturbation theory. Three
approaches have been discussed: in the ``exclusive approach'', all
$1/m_Q^2$ operators in the HQET are classified and their matrix
elements estimated, leading to~\cite{FaNe,TMann}
$\delta_{1/m^2}=-(3\pm 2)\%$; the ``inclusive approach'' has been
used to derive the bound $\delta_{1/m^2}<-3\%$, and to estimate
that~\cite{Shif}$^,$\footnote{This bound has been criticised in
Ref.~61.}
%\protect\citelow{Kapu}
$\delta_{1/m^2}=-(7\pm 3)\%$; the ``hybrid approach'' combines the
virtues of the former two to obtain a more restrictive lower bound on
$\delta_{1/m^2}$. This leads to~\cite{Vcbnew}
\begin{equation}
   \delta_{1/m^2} = - 0.055\pm 0.025 \,.
\end{equation}

Combining the above results, adding the theoretical errors linearly
to be conservative, gives
\begin{equation}\label{F1}
   {\cal F}(1) = 0.91\pm 0.03
\end{equation}
for the normalization of the hadronic form factor at zero recoil.
Thus, the corrections to the heavy-quark limit amount to a moderate
decrease of the form factor of about 10\%. This can be used to
extract from the experimental result (\ref{VcbF}) the
model-independent value
\begin{equation}\label{Vcbexc}
   |V_{cb}| = (38.8\pm 2.0_{\rm exp}\pm 1.2_{\rm th})
   \times 10^{-3} \,.
\end{equation}

\section{Heavy-Quark Effective Theory}

\subsection{The Effective Lagrangian}

The effects of a very heavy particle often become irrelevant at low
energies. It is then useful to construct a low-energy effective
theory, in which this heavy particle no longer appears~\cite{Anee}.
Eventually, this effective theory will be easier to deal with than
the full theory. A familiar example is Fermi's theory of the weak
interactions. For the description of weak decays of hadrons, the weak
interactions can be approximated by point-like four-fermion
couplings, governed by a dimensionful coupling constant $G_F$. Only
at energies much larger than the masses of hadrons can the effects of
the intermediate vector bosons, $W$ and $Z$, be resolved.

The process of removing the degrees of freedom of a heavy particle
involves the following steps~\cite{SVZ1}$^-$\cite{Polc}: one first
identifies the heavy-particle fields and ``integrates them out'' in
the generating functional of the Green functions of the theory. This
is possible since at low energies the heavy particle does not appear
as an external state. However, although the action of the full theory
is usually a local one, what results after this first step is a
non-local effective action. The non-locality is related to the fact
that in the full theory the heavy particle with mass $M$ can appear
in virtual processes and propagate over a short but finite distance
$\Delta x\sim 1/M$. Thus, a second step is required to obtain a local
effective Lagrangian: the non-local effective action is rewritten as
an infinite series of local terms using the OPE~\cite{Wils,Zimm}.
Roughly speaking, this corresponds to an expansion in powers of
$1/M$. It is in this step that the short- and long-distance physics
is disentangled. The long-distance physics corresponds to
interactions at low energies and is the same in the full and the
effective theory. But short-distance effects arising from quantum
corrections involving large virtual momenta (of order $M$) are not
reproduced in the effective theory, once the heavy particle has been
integrated out. In a third step, they have to be added in a
perturbative way using renormalization-group techniques. These
short-distance effects lead to a renormalization of the coefficients
of the local operators in the effective Lagrangian. An example is the
effective Lagrangian for non-leptonic weak decays, in which radiative
corrections from hard gluons with virtual momenta in the range
between $M_W$ and some renormalization scale $\mu\sim 1$~GeV give
rise to Wilson coefficients, which renormalize the local four-fermion
interactions~\cite{AltM}$^-$\cite{Gilm}.

\begin{figure}[htb]
   \epsfysize=8cm
   \centerline{\epsffile{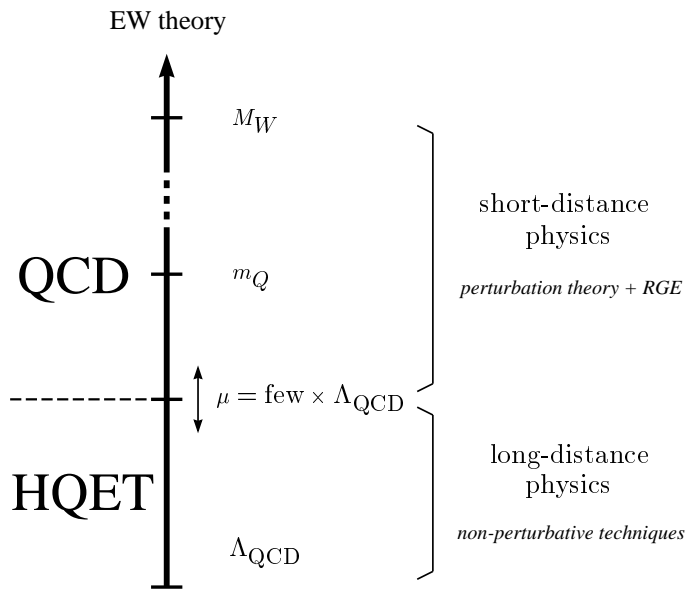}}
\caption{\label{fig:magic}
Philosophy of the heavy-quark effective theory.}
\end{figure}

The heavy-quark effective theory (HQET) is constructed to provide a
simplified description of processes where a heavy quark interacts
with light degrees of freedom predominantly by the exchange of soft
gluons~\cite{EiFe}$^-$\cite{Mann}. Clearly, $m_Q$ is the high-energy
scale in this case, and $\Lambda_{\rm QCD}$ is the scale of the
hadronic physics we are interested in. The situation is illustrated
in Fig.~\ref{fig:magic}. At short distances, i.e.\ for energy scales
larger than the heavy-quark mass, the physics is perturbative and
described by ordinary QCD. For mass scales much below the heavy-quark
mass, the physics is complicated and non-perturbative because of
confinement. Our goal is to obtain a simplified description in this
region using an effective field theory. To separate short- and
long-distance effects, we introduce a separation scale $\mu$ such
that $\Lambda_{\rm QCD}\ll\mu\ll m_Q$. The HQET will be constructed
in such a way that it is identical to QCD in the long-distance
region, i.e.\ for scales below $\mu$. In the short-distance region,
the effective theory is incomplete, however, since some high-momentum
modes have been integrated out from the full theory. The fact that
the physics must be independent of the arbitrary scale $\mu$ allows
us to derive renormalization-group equations, which can be employed
to deal with the short-distance effects in an efficient way.

Compared with most effective theories, in which the degrees of
freedom of a heavy particle are removed completely from the
low-energy theory, the HQET is special in that its purpose is to
describe the properties and decays of hadrons which do contain a
heavy quark. Hence, it is not possible to remove the heavy quark
completely from the effective theory. What is possible is to
integrate out the ``small components'' in the full heavy-quark
spinor, which describe the fluctuations around the mass shell.

The starting point in the construction of the low-energy effective
theory is the observation that a very heavy quark bound inside a
hadron moves more or less with the hadron's velocity $v$, and is
almost on shell. Its momentum can be written as
\begin{equation}\label{kresdef}
   p_Q^\mu = m_Q v^\mu + k^\mu \,,
\end{equation}
where the components of the so-called residual momentum $k$ are much
smaller than $m_Q$. Note that $v$ is a four-velocity, so that
$v^2=1$. Interactions of the heavy quark with light degrees of
freedom change the residual momentum by an amount of order $\Delta
k\sim\Lambda_{\rm QCD}$, but the corresponding changes in the
heavy-quark velocity vanish as $\Lambda_{\rm QCD}/m_Q\to 0$. In this
situation, it is appropriate to introduce large- and small-component
fields, $h_v$ and $H_v$, by
\begin{equation}\label{hvHvdef}
   h_v(x) = e^{i m_Q v\cdot x}\,P_+\,Q(x) \,, \qquad
   H_v(x) = e^{i m_Q v\cdot x}\,P_-\,Q(x) \,,
\end{equation}
where $P_+$ and $P_-$ are projection operators defined as
\begin{equation}
   P_\pm = {1\pm\rlap/v\over 2} \,.
\end{equation}
It follows that
\begin{equation}\label{redef}
   Q(x) = e^{-i m_Q v\cdot x}\,[ h_v(x) + H_v(x) ] \,.
\end{equation}
Because of the projection operators, the new fields satisfy
$\rlap/v\,h_v=h_v$ and $\rlap/v\,H_v=-H_v$. In the rest frame, i.e.\
for $v^\mu=(1,0,0,0)$, $h_v$ corresponds to the upper two components
of $Q$, while $H_v$ corresponds to the lower ones. Whereas $h_v$
annihilates a heavy quark with velocity $v$, $H_v$ creates a heavy
antiquark with velocity $v$.

\begin{figure}[htb]
   \epsfxsize=7cm
   \centerline{\epsffile{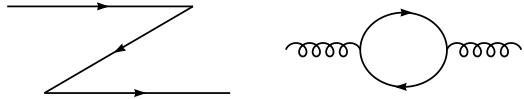}}
\caption{\label{fig:3.1}
Virtual fluctuations involving pair creation of heavy quarks. In the
first diagram, time flows to the right.}
\end{figure}

In terms of the new fields, the QCD Lagrangian for a heavy quark
takes the form
\begin{eqnarray}\label{Lhchi}
   {\cal L}_Q &=& \bar Q\,(i\,\rlap{\,/}D - m_Q)\,Q \nonumber\\
   &=& \bar h_v\,i v\!\cdot\!D\,h_v
    - \bar H_v\,(i v\!\cdot\!D + 2 m_Q)\,H_v \nonumber\\
   &&\mbox{}+ \bar h_v\,i\,\rlap{\,/}D_\perp H_v
    + \bar H_v\,i\,\rlap{\,/}D_\perp h_v \,,
\end{eqnarray}
where $D_\perp^\mu = D^\mu - v^\mu\,v\cdot D$ is orthogonal to the
heavy-quark velocity: $v\cdot D_\perp=0$. In the rest frame,
$D_\perp^\mu=(0,\vec D\,)$ contains the spatial components of the
covariant derivative. From (\ref{Lhchi}), it is apparent that $h_v$
describes massless degrees of freedom, whereas $H_v$ corresponds to
fluctuations with twice the heavy-quark mass. These are the heavy
degrees of freedom that will be eliminated in the construction of the
effective theory. The fields are mixed by the presence of the third
and fourth terms, which describe pair creation or annihilation of
heavy quarks and antiquarks. As shown in the first diagram in
Fig.~\ref{fig:3.1}, in a virtual process a heavy quark propagating
forward in time can turn into an antiquark propagating backward in
time, and then turn back into a quark. The energy of the intermediate
quantum state $h h\bar H$ is larger than the energy of the initial
heavy quark by at least $2 m_Q$. Because of this large energy gap,
the virtual quantum fluctuation can only propagate over a short
distance $\Delta x\sim 1/m_Q$. On hadronic scales set by $R_{\rm
had}\sim 1/\Lambda_{\rm QCD}$, the process essentially looks like a
local interaction of the form
\begin{equation}
   \bar h_v\,i\,\rlap{\,/}D_\perp\,{1\over 2 m_Q}\,
   i\,\rlap{\,/}D_\perp h_v \,,
\end{equation}
where we have simply replaced the propagator for $H_v$ by $1/2 m_Q$.
A more correct treatment is to integrate out the small-component
field $H_v$, thereby deriving a non-local effective action for the
large-component field $h_v$, which can then be expanded in terms of
local operators. Before doing this, let us mention a second type of
virtual corrections involving pair creation, namely heavy-quark
loops. An example is shown in the second diagram in
Fig.~\ref{fig:3.1}. Heavy-quark loops cannot be described in terms of
the effective fields $h_v$ and $H_v$, since the quark velocities
inside a loop are not conserved and are in no way related to hadron
velocities. However, such short-distance processes are proportional
to the small coupling constant $\alpha_s(m_Q)$ and can be calculated
in perturbation theory. They lead to corrections that are added onto
the low-energy effective theory in the renormalization procedure.

On a classical level, the heavy degrees of freedom represented by
$H_v$ can be eliminated using the equation of motion. Taking the
variation of the Lagrangian with respect to the field $\bar H_v$, we
obtain
\begin{equation}
   (i v\!\cdot\!D + 2 m_Q)\,H_v = i\,\rlap{\,/}D_\perp h_v \,.
\end{equation}
This equation can formally be solved to give
\begin{equation}\label{Hfield}
   H_v = {1\over 2 m_Q + i v\!\cdot\!D}\,
   i\,\rlap{\,/}D_\perp h_v \,,
\end{equation}
showing that the small-component field $H_v$ is indeed of order
$1/m_Q$. We can now insert this solution into (\ref{Lhchi}) to obtain
the ``non-local effective Lagrangian''
\begin{equation}\label{Lnonloc}
   {\cal L}_{\rm eff} = \bar h_v\,i v\!\cdot\!D\,h_v
   + \bar h_v\,i\,\rlap{\,/}D_\perp\,{1\over 2 m_Q+i v\!\cdot\!D}\,
   i\,\rlap{\,/}D_\perp h_v \,.
\end{equation}
Clearly, the second term corresponds to the first class of virtual
processes shown in Fig.~\ref{fig:3.1}.

It is possible to derive this Lagrangian in a more elegant way by
manipulating the generating functional for QCD Green's functions
containing heavy-quark fields~\cite{Mann}. To this end, one starts
from the field redefinition (\ref{redef}) and couples the
large-component fields $h_v$ to external sources $\rho_v$. Green's
functions with an arbitrary number of $h_v$ fields can be constructed
by taking derivatives with respect to $\rho_v$. No sources are needed
for the heavy degrees of freedom represented by $H_v$. The functional
integral over these fields is Gaussian and can be performed
explicitly, leading to the effective action
\begin{equation}\label{SeffMRR}
   S_{\rm eff} = \int\!{\rm d}^4 x\,{\cal L}_{\rm eff}
   - i \ln\Delta \,,
\end{equation}
with ${\cal L}_{\rm eff}$ as given in (\ref{Lnonloc}). The
appearance
of the logarithm of the determinant
\begin{equation}
   \Delta = \exp\bigg( {1\over 2}\,{\rm Tr}\,
   \ln\big[ 2 m_Q + i v\!\cdot\!D - i\eta \big] \bigg)
\end{equation}
is a quantum effect not present in the classical derivation presented
above. However, in this case the determinant can be regulated in a
gauge-invariant way, and by choosing the gauge $v\cdot A=0$ one can
show that $\ln\Delta$ is just an irrelevant
constant~\cite{Mann,Soto}.

\begin{figure}[htb]
   \epsfysize=3cm
   \centerline{\epsffile{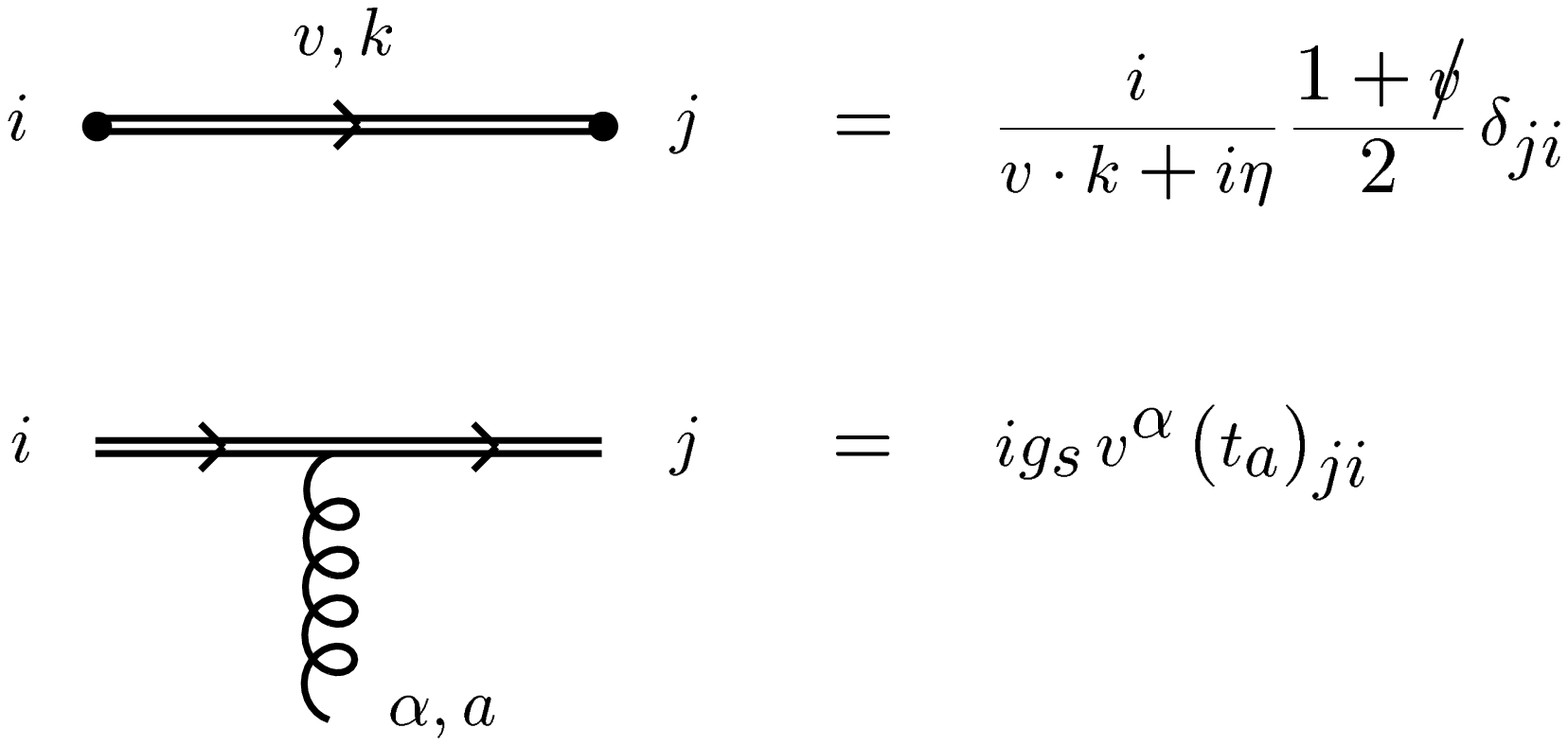}}
\caption{\label{fig:3.2}
Feynman rules of the HQET ($i,j$ and $a$ are colour indices). A heavy
quark is represented by a double line labelled by the velocity $v$
and the residual momentum $k$. The velocity $v$ is conserved by the
strong interactions.}
\end{figure}

Because of the phase factor in (\ref{redef}), the $x$ dependence of
the effective heavy-quark field $h_v$ is weak. In momentum space,
derivatives acting on $h_v$ correspond to powers of the residual
momentum $k$, which by construction is much smaller than $m_Q$.
Hence, the non-local effective Lagrangian (\ref{Lnonloc}) allows for
a derivative expansion in powers of $iD/m_Q$:
\begin{equation}
   {\cal L}_{\rm eff} = \bar h_v\,i v\!\cdot\!D\,h_v
   + {1\over 2 m_Q}\,\sum_{n=0}^\infty\,
   \bar h_v\,i\,\rlap{\,/}D_\perp\,\bigg( -{i v\cdot D\over 2 m_Q}
   \bigg)^n\,i\,\rlap{\,/}D_\perp h_v \,.
\end{equation}
Taking into account that $h_v$ contains a $P_+$ projection operator,
and using the identity
\begin{equation}\label{pplusid}
   P_+\,i\,\rlap{\,/}D_\perp\,i\,\rlap{\,/}D_\perp P_+
   = P_+\,\bigg[ (i D_\perp)^2 + {g_s\over 2}\,
   \sigma_{\mu\nu }\,G^{\mu\nu } \bigg]\,P_+ \,,
\end{equation}
where $[i D^\mu,i D^\nu]=i g_s G^{\mu\nu}$ is the gluon
field-strength tensor, one finds that~\cite{EiH2,FGL}
\begin{equation}\label{Lsubl}
   {\cal L}_{\rm eff} = \bar h_v\,i v\!\cdot\!D\,h_v
   + {1\over 2 m_Q}\,\bar h_v\,(i D_\perp)^2\,h_v
   + {g_s\over 4 m_Q}\,\bar h_v\,\sigma_{\mu\nu}\,
   G^{\mu\nu}\,h_v + O(1/m_Q^2) \,.
\end{equation}
In the limit $m_Q\to\infty$, only the first terms remains:
\begin{equation}\label{Leff}
   {\cal L}_\infty = \bar h_v\,i v\!\cdot\!D\,h_v \,.
\end{equation}
This is the effective Lagrangian of the HQET. It gives rise to the
Feynman rules depicted in Fig.~\ref{fig:3.2}.

Let us take a moment to study the symmetries of this
Lagrangian~\cite{Geor}. Since there appear no Dirac matrices,
interactions of the heavy quark with gluons leave its spin unchanged.
Associated with this is an SU(2) symmetry group, under which
${\cal L}_\infty$ is invariant. The action of this symmetry on the
heavy-quark fields becomes most transparent in the rest frame, where
the generators $S^i$ of SU(2) can be chosen as
\begin{equation}\label{Si}
   S^i = {1\over 2} \left( \begin{array}{cc}
                           \sigma^i ~&~ 0 \\
                           0 ~&~ \sigma^i \end{array} \right) \,,
   \qquad [S^i,S^j] = i \epsilon^{ijk} S^k \,.
\end{equation}
Here $\sigma^i$ are the Pauli matrices. An infinitesimal SU(2)
transformation $h_v\to (1 + i\vec\epsilon \cdot\vec S\,)\,h_v$ leaves
the Lagrangian invariant:
\begin{equation}\label{SU2tr}
   \delta{\cal L}_\infty = \bar h_v\,
   [i v\!\cdot\! D,i \vec\epsilon\cdot\vec S\,]\,h_v = 0 \,.
\end{equation}
Another symmetry of the HQET arises since the mass of the heavy quark
does not appear in the effective Lagrangian. For $N_h$ heavy quarks
moving at the same velocity, eq.~(\ref{Leff}) can be extended by
writing
\begin{equation}\label{Leff2}
   {\cal L}_\infty
   = \sum_{i=1}^{N_h}\,\bar h_v^i\,i v\!\cdot\! D\,h_v^i \,.
\end{equation}
This is invariant under rotations in flavour space. When combined
with the spin symmetry, the symmetry group is promoted to SU$(2N_h)$.
This is the heavy-quark spin--flavour symmetry~\cite{Isgu,Geor}. Its
physical content is that, in the limit $m_Q\to\infty$, the strong
interactions of a heavy quark become independent of its mass and
spin.

Consider now the operators appearing at order $1/m_Q$ in the
effective Lagrangian (\ref{Lsubl}). They are easiest to identify in
the rest frame. The first operator,
\begin{equation}\label{Okin}
   {\cal O}_{\rm kin} = {1\over 2 m_Q}\,\bar h_v\,(i D_\perp)^2\,
   h_v \to - {1\over 2 m_Q}\,\bar h_v\,(i \vec D\,)^2\,h_v \,,
\end{equation}
is the gauge-covariant extension of the kinetic energy arising from
the off-shell residual motion of the heavy quark. The second operator
is the non-abelian analogue of the Pauli interaction, which describes
the chromo-magnetic coupling of the heavy-quark spin to the gluon
field:
\begin{equation}\label{Omag}
   {\cal O}_{\rm mag} = {g_s\over 4 m_Q}\,\bar h_v\,
   \sigma_{\mu\nu}\,G^{\mu\nu}\,h_v \to
   - {g_s\over m_Q}\,\bar h_v\,\vec S\!\cdot\!\vec B_c\,h_v \,.
\end{equation}
Here $\vec S$ is the spin operator defined in (\ref{Si}), and $B_c^i
= -\frac{1}{2}\epsilon^{ijk} G^{jk}$ are the components of the
chromo-magnetic field. The chromo-magnetic interaction is a
relativistic effect, which scales like $1/m_Q$. This is the origin of
the heavy-quark spin symmetry.

\subsection{The Residual Mass Term and the Definition of the
Heavy-Quark Mass}

The choice of the expansion parameter in the HQET, i.e.\ the
definition of the heavy-quark mass $m_Q$, deserves some comments. In
the derivation presented earlier in this section, we chose $m_Q$ to
be the ``mass in the Lagrangian'', and using this parameter in the
phase redefinition in (\ref{redef}) we obtained the effective
Lagrangian (\ref{Leff}), in which the heavy-quark mass no longer
appears. However, this treatment has its subtleties. The symmetries
of the HQET allow a ``residual mass'' $\delta m$ for the heavy quark,
provided that $\delta m$ is of order $\Lambda_{\rm QCD}$ and is the
same for all heavy-quark flavours. Even if we arrange that such a
mass term is not present at the tree level, it will in general be
induced by quantum corrections. (This is unavoidable if the theory is
regulated with a dimensionful cutoff.) Therefore, instead of
(\ref{Leff}) we should write the effective Lagrangian in the more
general form~\cite{FNL}:
\begin{eqnarray}
   h_v(x) &=& e^{i m_Q v\cdot x}\,P_+\,Q(x) \nonumber\\
   \Rightarrow \qquad
   {\cal L}_\infty &=& \bar h_v\,iv\!\cdot\!D\,h_v
    - \delta m\,\bar h_v h_v \,.
\end{eqnarray}
If we redefine the expansion parameter according to $m_Q\to
m_Q+\Delta m$, the residual mass changes in the opposite way: $\delta
m\to\delta m-\Delta m$. This implies that there is a unique choice of
the expansion parameter $m_Q$ such that $\delta m=0$. Requiring
$\delta m=0$, as it is usually done implicitly in the HQET, defines a
heavy-quark mass, which in perturbation theory coincides with the
pole mass~\cite{Tarr}. This, in turn, defines for each heavy hadron a
parameter $\bar\Lambda$ (sometimes called the ``binding energy'')
through
\begin{equation}
   \bar\Lambda = (m_H - m_Q)\Big|_{m_Q\to\infty} \,.
\end{equation}
If one prefers to work with another choice of the expansion
parameter, the values of non-perturbative parameters such as
$\bar\Lambda$ change, but at the same time one has to include the
residual mass term in the HQET Lagrangian. It can be shown that the
various parameters that depend on the definition of $m_Q$ enter the
predictions for all physical observables in such a way that the
results are independent of which particular choice one
adopts~\cite{FNL}.
       
There is one more subtlety hidden in the above discussion. The
quantities $m_Q$, $\bar\Lambda$ and $\delta m$ are non-perturbative
parameters of the HQET, which have a similar status as the vacuum
condensates in QCD phenomenology~\cite{SVZ}. These parameters cannot
be defined unambiguously in perturbation theory. The reason lies in 
the divergent behaviour of perturbative expansions in large orders,
which is associated with the existence of singularities along the
real axis in the Borel plane, the so-called
renormalons~\cite{tHof}$^-$\cite{Muel}. For instance, the
perturbation series which relates the pole mass $m_Q$ of a heavy
quark to its bare mass,
\begin{equation}
   m_Q = m_Q^{\rm bare}\,\Big\{ 1 + c_1\,\alpha_s(m_Q)
   + c_2\,\alpha_s^2(m_Q) + \dots + c_n\,\alpha_s^n(m_Q)
   + \dots \Big\} \,,
\end{equation}
contains numerical coefficients $c_n$ that grow as $n!$ for large
$n$, rendering the series divergent and not Borel
summable~\cite{BBren,Bigiren}. The best one can achieve is to
truncate the perturbation series at the minimal term, but this leads
to an unavoidable arbitrariness of order $\Delta m_Q\sim\Lambda_{\rm
QCD}$ (the size of the minimal term). This observation, which at
first sight seems a serious problem for QCD phenomenology, should
actually not come as a surprise. We know that because of confinement
quarks do not appear as physical states in nature. Hence, there is no
unique way to define their on-shell properties such as a pole mass.
In view of this, it is actually remarkable that QCD perturbation
theory ``knows'' about its incompleteness and indicates, through the
appearance of renormalon singularities, the presence of
non-perturbative effects. We must first specify a scheme how to
truncate the QCD perturbation series before non-perturbative
statements such as $\delta m=0$ become meaningful, and hence before
non-perturbative parameters such as $m_Q$ and $\bar\Lambda$ become
well-defined quantities. The actual values of these parameters will
depend on this scheme.

We stress that the ``renormalon ambiguities'' are not a conceptual
problem for the heavy-quark expansion. In fact, it can be shown quite
generally that these ambiguities cancel in all predictions for
physical observables~\cite{Chris}$^-$\cite{LMS}. The way the
cancellations occur is intricate, however. The generic structure of
the heavy-quark expansion for an observable is of the form:
\begin{equation}
   \mbox{observable} \sim C[\alpha_s(m_Q)]\,\bigg( 1
   + {\Lambda\over m_Q} + \dots \bigg) \,.
\end{equation}
Here $C[\alpha_s(m_Q)]$ represents a perturbative coefficient
function, and $\Lambda$ is a dimensionful non-perturbative parameter.
The truncation of the perturbation series defining the coefficient
function leads to an arbitrariness of order $\Lambda_{\rm QCD}/m_Q$,
which precisely cancels against a corresponding arbitrariness of
order $\Lambda_{\rm QCD}$ in the definition of the non-perturbative
parameter $\Lambda$.

The renormalon problem poses itself when one imagines to apply
perturbation theory in very high orders. In practise, the
perturbative coefficients are known to finite order in $\alpha_s$ (at
best to two-loop accuracy), and to be consistent one should use them
in connection with the pole mass (and $\bar\Lambda$ etc.) defined to
the same order. For completeness, we note that for the ``one-loop
pole masses'' of the heavy quarks we shall adopt the values
\begin{equation}\label{masses}
   m_b = 4.8~\mbox{GeV} \,,\qquad m_c = 1.4~\mbox{GeV} \,.
\end{equation}
Their difference satisfies the constraint in (\ref{mbmcval}).

\section{Inclusive Decay Rates}
 
Inclusive decay rates determine the probability of the decay of a
particle into the sum of all possible final states with a given set
of quantum numbers. An example is provided by the inclusive
semileptonic decay rate of the $B$ meson, $\Gamma(\bar B\to
X_c\,\ell\,\bar\nu)$, where the final state consists of a
lepton--neutrino pair accompanied by any number of hadrons with total
charm-quark number $n_c=1$. Here we shall discuss the
theoretical description of inclusive decays of hadrons containing a
heavy quark~\cite{Chay}$^-$\cite{Fermi}. From the theoretical point
of view, such decays have two advantages: first, bound-state effects
related to the initial state (such as the ``Fermi motion'' of the
heavy quark inside the hadron) can be accounted for in a systematic
way using the heavy-quark expansion, in much the same way as
explained in the previous sections; secondly, the fact that the final
state consists of a sum over many hadronic channels eliminates
bound-state effects related to the properties of individual hadrons.
This second feature is based on a hypothesis known as quark--hadron
duality, which is an important concept in QCD phenomenology. The
assumption of duality is that cross sections and decay rates, which
are defined in the physical region (i.e.\ the region of time-like
momenta), are calculable in QCD after a ``smearing'' or ``averaging''
procedure has been applied~\cite{PQW}. In semileptonic decays, it is
the integration over the lepton and neutrino phase space that
provides a ``smearing'' over the invariant hadronic mass of the final
state (so-called ``global'' duality). For non-leptonic decays, on the
other hand, the total hadronic mass is fixed, and it is only the fact
that one sums over many hadronic states that provides an
``averaging'' (so-called ``local'' duality). Clearly, local duality
is a stronger assumption than global duality. It is important to
stress that quark--hadron duality cannot yet be derived from first
principles, although it is a necessary assumption for many
applications of QCD. The validity of global duality has been tested
experimentally using data on hadronic $\tau$ decays~\cite{Maria}. A
more formal attempt to address the problem of quark--hadron duality
can be found in Ref.~108.
%\citelow{Shifm}

Using the optical theorem, the inclusive decay width of a hadron
$H_b$ containing a $b$ quark can be written in the form
\begin{equation}\label{ImT}
   \Gamma(H_b\to X) = {1\over 2 m_{H_b}}\,2\,\mbox{Im}\,
   \langle H_b|\,{\bf T}\,|H_b\rangle \,,
\end{equation}
where the transition operator ${\bf T}$ is given by
\begin{equation}
   {\bf T} = i\!\int{\rm d}^4x\,T\{\,
   {\cal L}_{\rm eff}(x),{\cal L}_{\rm eff}(0)\,\} \,.
\end{equation}
In fact, inserting a complete set of states inside the time-ordered
product, we recover the standard expression
\begin{equation}
   \Gamma(H_b\to X) = {1\over 2 m_{H_b}}\,\sum_X\,
   (2\pi)^4\,\delta^4(p_H-p_X)\,|\langle X|\,{\cal L}_{\rm eff}\,
   |H_b\rangle|^2
\end{equation}
for the decay rate. Here ${\cal L}_{\rm eff}$ is the effective weak
Lagrangian corrected for short-distance
effects~\cite{AltM,Gail,cpcm3}$^-$\cite{cpcm5} arising from the
exchange of gluons with virtualities between $m_W$ and $m_b$. If some
quantum numbers of the final states $X$ are specified, the sum over
intermediate states is to be restricted appropriately. In the case of
the inclusive semileptonic decay rate, for instance, the sum would
include only those states $X$ containing a lepton--neutrino pair.

\begin{figure}[htb]
   \epsfxsize=7cm
   \centerline{\epsffile{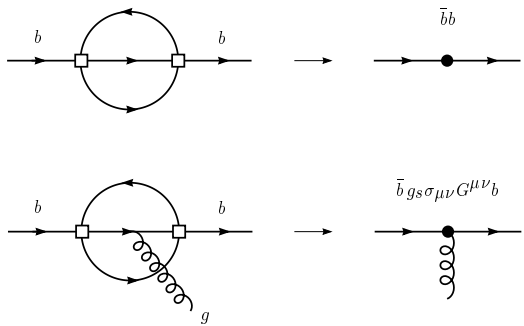}}
\caption{\label{fig:Toper}
Perturbative contributions to the transition operator ${\bf T}$
(left), and the corresponding operators in the OPE (right). The open
squares represent a four-fermion interaction of the effective
Lagrangian ${\cal L}_{\rm eff}$, while the black circles represent
local operators in the OPE.}
\end{figure}

In perturbation theory, some contributions to the transition operator
are given by the two-loop diagrams shown on the left-hand side in
Fig.~\ref{fig:Toper}. Because of the large mass of the $b$ quark, the
momenta flowing through the internal lines in these diagrams are
large. It is thus possible to construct an OPE for the transition
operator, in which ${\bf T}$ is represented as a series of local
operators containing the heavy-quark fields. The operator with the
lowest dimension, $d=3$, is $\bar b b$. It arises from integrating
over the internal lines in the first diagram shown in the figure. The
only gauge-invariant operator with dimension $d=4$ is $\bar
b\,i\rlap{\,/}D\,b$; however, the equation of motion implies that
between physical states this operator can be replaced by $m_b\bar b
b$. The first operator that is different from $\bar b b$ has
dimension $d=5$ and contains the gluon field. It is given by $\bar
b\,g_s\sigma_{\mu\nu} G^{\mu\nu} b$. This operator arises from
diagrams in which a gluon is emitted from one of the internal lines,
such as the second diagram shown in the figure.

For dimensional reasons, the matrix elements of such
higher-dimensional operators are suppressed by inverse powers of the
heavy-quark mass. Thus, any inclusive decay rate of a hadron $H_b$
can be written in the form~\cite{Bigi}$^-$\cite{MaWe}:
\begin{equation}\label{gener}
   \Gamma(H_b\to X_f) = {G_F^2 m_b^5\over 192\pi^3}\,
   \bigg\{ c_3^f\,\langle\bar b b\rangle_H
   + c_5^f\,{\langle\bar b\,g_s\sigma_{\mu\nu} G^{\mu\nu} b
   \rangle_H\over m_b^2} + \dots \bigg\} \,,
\end{equation}
where the prefactor arises naturally from the loop integrations,
$c_n^f$ are calculable coefficient functions (which also contain the
relevant CKM matrix elements) depending on the quantum numbers $f$ of
the final state, and $\langle O\rangle_H$ are the (normalized)
forward matrix elements of local operators, for which we use the
short-hand notation
\begin{equation}
   \langle O\rangle_H = {1\over 2 m_{H_b}}\,\langle H_b|\,
   O\,|H_b\rangle \,.
\end{equation}

In the next step, these matrix elements are systematically expanded
in powers of $1/m_b$, using the technology of the HQET. Introducing
the velocity-dependent fields $b_v$ of the HQET, where $v$ denotes
the velocity of the hadron $H_b$, one finds~\cite{FaNe,MaWe,Adam}
\begin{eqnarray}
   \langle\bar b b\rangle_H &=& 1
    - {\mu_\pi^2(H_b)-\mu_G^2(H_b)\over 2 m_b^2} + O(1/m_b^3) \,,
    \nonumber\\
   \langle\bar b\,g_s\sigma_{\mu\nu} G^{\mu\nu} b\rangle_H
   &=& 2\mu_G^2(H_b) + O(1/m_b) \,,
\end{eqnarray}
where we have defined the HQET matrix elements
\begin{eqnarray}
   \mu_\pi^2(H_b) &=& {1\over 2 m_{H_b}}\,
    \langle H_b(v)|\,\bar b_v\,(i\vec D)^2\,b_v\,|H_b(v)\rangle \,,
    \nonumber\\
   \mu_G^2(H_b) &=& {1\over 2 m_{H_b}}\,
    \langle H_b(v)|\,\bar b_v {g_s\over 2}\sigma_{\mu\nu}
     G^{\mu\nu} b_v\,|H_b(v)\rangle \,.
\end{eqnarray}
Here $(i\vec D)^2=(i v\cdot D)^2-(i D)^2$; in the rest frame, this is
the square of the operator for the
spatial momentum of the heavy quark. Inserting these results into
(\ref{gener}), we obtain
\begin{equation}\label{generic}
   \Gamma(H_b\to X_f) = {G_F^2 m_b^5\over 192\pi^3}\,
   \bigg\{ c_3^f\,\bigg( 1 - {\mu_\pi^2(H_b)\over 2 m_b^2} \bigg)
   + (4 c_5^f + c_3^f)\,{\mu_G^2(H_b)\over 2 m_b^2}
   + \dots \bigg\} \,.
\end{equation}
It is instructive to understand the appearance of the ``kinetic
energy'' contribution $\mu_\pi^2$, which is the gauge-covariant
extension of the square of the $b$-quark momentum inside the heavy
hadron. This contribution is the field-theory analogue of the Lorentz
factor $(1-\vec v_b^{\,2})^{1/2}\simeq 1-\vec p_b^{\,2}/2 m_b^2$, in
accordance with the fact that the lifetime, $\tau=1/\Gamma$, for a
moving particle increases due to time dilation.

The main result of the heavy-quark expansion for inclusive decay
rates is that the free quark decay (i.e.\ the parton model) provides
the first term in a systematic $1/m_b$ expansion, i.e.\
\begin{equation}
   \Gamma(H_b\to X_f) = {G_F^2 m_b^5\over 192\pi^3}\,c_3^f\,
   \Big\{ 1 + O(1/m_b^2) \Big\} \,.
\end{equation}
For dimensional reasons, the free-quark decay rate is proportional to
the fifth power of the $b$-quark mass. The non-perturbative
corrections to this picture, which arise from bound-state effects
inside the hadron $H_b$, are suppressed by (at least) two powers of
the heavy-quark mass, i.e.\ they are of relative order $(\Lambda_{\rm
QCD}/m_b)^2$. Note that the absence of first-order power corrections
is a simple consequence of the equation of motion, as there is no
independent gauge-invariant operator of dimension $d=4$ that could
appear in the OPE. The fact that bound-state effects in inclusive
decays are strongly suppressed explains a posteriori the success of
the parton model in describing such processes.

The hadronic matrix elements appearing in the heavy-quark expansion
(\ref{generic}) can be determined to some extent from the known
masses of heavy hadron states. For the $B$ meson, one finds
that~\cite{FaNe}
\begin{eqnarray}\label{mupimuG}
   \mu_\pi^2(B) &=& - \lambda_1 = (0.3\pm 0.2)~\mbox{GeV}^2 \,,
    \nonumber\\
   \mu_G^2(B) &=& 3\lambda_2 = {3\over 4}\,(m_{B^*}^2 - m_B^2)
    \simeq 0.36~\mbox{GeV}^2 \,,
\end{eqnarray}
where $\lambda_1$ and $\lambda_2$ are the parameters appearing in the
mass formula (\ref{FNrela}), and their numerical values have been
taken from (\ref{lam1}) and (\ref{lam2val}). For the ground-state
baryon $\Lambda_b$, in which the light constituents have total spin
zero, it follows that
\begin{equation}
   \mu_G^2(\Lambda_b) = 0 \,,
\end{equation}
while the matrix element $\mu_\pi^2(\Lambda_b)$ obeys the relation
\begin{equation}
   (m_{\Lambda_b}-m_{\Lambda_c}) - (\overline{m}_B-\overline{m}_D)
   = \Big[ \mu_\pi^2(B)-\mu_\pi^2(\Lambda_b) \Big]\,\bigg(
   {1\over 2 m_c} - {1\over 2 m_b} \bigg) + O(1/m_Q^2) \,,
\end{equation}
where $\overline{m}_B$ and $\overline{m}_D$ denote the spin-averaged
masses introduced in connection with (\ref{mbmc}). With the value of
$m_{\Lambda_b}$ given in (\ref{Lbmass}), this leads to
\begin{equation}\label{mupidif}
   \mu_\pi^2(B) - \mu_\pi^2(\Lambda_b) = (0.01\pm 0.03)~\mbox{GeV}^2
   \,.
\end{equation}
What remains to be calculated, then, is the coefficient functions
$c_n^f$ for a given inclusive decay channel. We shall now discuss two
important applications of this general formalism.

\subsection{Determination of $|V_{cb}|$ from Inclusive Semileptonic
Decays}
 
The extraction of $|V_{cb}|$ from the inclusive semileptonic decay
rate of the $B$ meson is based on the
expression~\cite{Bigi}$^-$\cite{MaWe}
\begin{eqnarray}\label{Gamsl}
   \Gamma(\bar B\to X_c\,\ell\,\bar\nu)
   &=& {G_F^2 m_b^5\over 192\pi^3}\,|V_{cb}|^2\,\Bigg\{\!\bigg(
    1 + {\lambda_1+3\lambda_2\over 2 m_b^2} \bigg)\!\bigg[
    f\bigg( {m_c\over m_b} \bigg)
    + {\alpha_s(M)\over\pi}\,g\bigg( {m_c\over m_b} \bigg) \bigg]
    \nonumber\\
   &&\phantom{ {G_F^2 m_b^5\over 192\pi^3}\,|V_{cb}|^2\,\Bigg\{ }
    - {6\lambda_2\over m_b^2}\,\bigg( 1 - {m_c^2\over m_b^2}
    \bigg)^4  + \dots \Bigg\} \,,
\end{eqnarray}
where $m_b$ and $m_c$ are the poles mass of the $b$ and $c$ quarks
(defined to a given order in perturbation theory), and $f(x)$ and
$g(x)$ are phase-space functions:
\begin{equation}
   f(x) = 1 - 8 x^2 + 8 x^6 - x^8 - 12 x^4\ln x^2 \,,
\end{equation}
and $g(x)$ is given elsewhere~\cite{fgrefs}. The theoretical
uncertainties in this determination of $|V_{cb}|$ are quite
different from those entering the analysis of exclusive decays. In
particular, in inclusive decays there appear the quark masses rather
than the meson masses. Moreover, the theoretical description relies
on the assumption of global quark--hadron duality, which is not
necessary for exclusive decays. 

A careful analysis shows that the main sources of theoretical
uncertainties are the dependence on the heavy-quark masses and
unknown higher-order perturbative corrections~\cite{Beijing}. Each
lead to an uncertainty of $\delta\Gamma/\Gamma\simeq 10\%$ in the
prediction for the semileptonic decay rate. Adding, as previously,
the theoretical errors linearly, and taking the square root, leads to
\begin{equation} {\delta|V_{cb}|\over|V_{cb}|} \simeq 10\%
\end{equation} for the theoretical uncertainty in the determination
of $|V_{cb}|$ from inclusive decays, keeping in mind that this method
relies in addition on the assumption of global quark--hadron duality.
Taking the result of Ball et al.\ for the central value~\cite{BaBB},
we quote
\begin{equation}
   |V_{cb}| = (0.0400\pm 0.0040)\,\bigg( {B_{\rm SL}\over 10.9\%}
   \bigg)^{1/2}\,\bigg( {\tau_B\over 1.6~\mbox{ps}}
   \bigg)^{-1/2} \,.
\end{equation}
With the new world averages for the semileptonic branching ratio,
$B_{\rm SL}=(10.90\pm 0.46)\%$ (see below), and for the average
$B$-meson lifetime~\cite{Joe}, $\tau_B=(1.60\pm 0.03)$~ps, we
obtain
\begin{equation}
   |V_{cb}| = (40.0\pm 0.9_{\rm exp}\pm 4.0_{\rm th})
   \times 10^{-3} \,.
\end{equation}
This is in excellent agreement with the value in (\ref{Vcbexc}),
which has been extracted from the analysis of the exclusive decay
$\bar B\to D^*\ell\,\bar\nu$. This agreement is gratifying given the
differences of the methods used, and it provides an indirect test of
global quark--hadron duality. Combining the two measurements gives
the final result
\begin{equation}
   |V_{cb}| = 0.039\pm 0.002 \,.
\end{equation}
After $V_{ud}$ and $V_{us}$, this is now the third-best known entry
in the CKM matrix.

\subsection{Semileptonic Branching Ratio and Charm Counting}
 
The semileptonic branching ratio of $B$ mesons is defined as
\begin{equation}
   B_{\rm SL} = {\Gamma(\bar B\to X\,e\,\bar\nu)\over
   \sum_\ell \Gamma(\bar B\to X\,\ell\,\bar\nu) + \Gamma_{\rm NL}
   + \Gamma_{\rm rare}} \,,
\end{equation}
where $\Gamma_{\rm NL}$ and $\Gamma_{\rm rare}$ are the inclusive
rates for non-leptonic and rare decays, respectively. The main
difficulty in calculating $B_{\rm SL}$ is not in the semileptonic
width, but in the non-leptonic one. As mentioned above, the
calculation of non-leptonic decays in the heavy-quark expansion
relies on the strong assumption of local quark--hadron duality.

Measurements of the semileptonic branching ratio have been performed
by various experimental groups, using both model-dependent and
model-in\-de\-pen\-dent analyses~\cite{Hons}. The status of the
results is controversial, as there is a discrepancy between
low-energy measurements performed at the $\Upsilon(4s)$ resonance and
high-energy measurements performed at the $Z^0$ resonance. The
average value at low energies is~\cite{Tomasz} $B_{\rm SL}=(10.37\pm
0.30)\%$. High-energy measurements performed at LEP, on the other
hand, give~\cite{Pascal} $B_{\rm SL}^{(b)}=(11.11\pm 0.23)\%$. The
superscript $(b)$ indicates that this value refers not to the $B$
meson, but to a mixture of $b$ hadrons (approximately 40\% $B^-$,
40\% $\bar B^0$, 12\% $B_s$, and 8\% $\Lambda_b$). Assuming that the
corresponding semileptonic width $\Gamma_{\rm SL}^{(b)}$ is close to
that of the $B$ meson,\footnote{Theoretically, this is expected to be
a very good approximation.}
we can correct for this and find $B_{\rm SL}=(\tau(B)/\tau(b))\,
B_{\rm SL}^{(b)}=(11.30\pm 0.26)\%$, where $\tau(b)=(1.57\pm
0.03)$~ps is the average lifetime corresponding to the above mixture
of $b$ hadrons~\cite{Joe}. The discrepancy between the low- and
high-energy measurements of the semileptonic branching ratio is
therefore larger than three standard deviations. If we take the
average and inflate the error to account for this fact, we obtain
\begin{equation}\label{Bslval}
   B_{\rm SL} = (10.90\pm 0.46)\% \,.
\end{equation}
In understanding this result, an important aspect is charm counting,
i.e.\ the measurement of the average number $n_c$ of charm hadrons
produced per $B$ decay. Theoretically, this quantity is given by
\begin{equation}\label{ncdef}
   n_c = 1 + B(\bar B\to X_{c\bar c s'})
   - B(\bar B\to\mbox{no charm}) \,,
\end{equation}
where $B(\bar B\to X_{c\bar c s'})$ is the branching ratio for decays
into final states containing two charm quarks, and $B(\bar
B\to\mbox{no charm})\sim 0.02$ is the Standard Model branching ratio
for charmless decays~\cite{Alta}$^-$\cite{Buch}. Recently, two new
measurements of the average charm content have been performed. The
CLEO Collaboration has presented the value~\cite{Tomasz,ncnew}
$n_c=1.16\pm 0.05$, and the ALEPH Collaboration has reported the
result~\cite{ALEPHnc} $n_c=1.23\pm 0.07$. The average is
\begin{equation}
   n_c = 1.18\pm 0.04 \,.
\end{equation}

The naive parton model predicts that $B_{\rm SL}\simeq 15\%$ and
$n_c\simeq 1.2$; however, it has been known for some time that
perturbative corrections could change these predictions
significantly~\cite{Alta}. With the establishment of the $1/m_Q$
expansion, the non-perturbative corrections to the parton model could
be computed, and their effect turned out to be very small. This led
Bigi et al.\ to conclude that values $B_{\rm SL}<12.5\%$ cannot be
accommodated by theory, thus giving rise to a puzzle referred to as
the ``baffling semileptonic branching ratio''~\cite{baff}. Later,
Bagan et al.\ have completed the calculation of the $O(\alpha_s)$
corrections including the effects of the charm-quark mass, finding
that they lower the value of $B_{\rm SL}$
significantly~\cite{BSLnew1}.

The original analysis of Bagan et al.\ has recently been corrected in
an erratum~\cite{BSLnew1}. Here we shall present the results of an
independent numerical analysis using the same theoretical input (for
a detailed discussion, see Ref.~124).
%\citelow{MNChris}
The semileptonic branching ratio and $n_c$ depend on the quark-mass
ratio $m_c/m_b$ and on the ratio $\mu/m_b$, where $\mu$ is the scale
used to renormalize the coupling constant $\alpha_s(\mu)$ and the
Wilson coefficients appearing in the non-leptonic decay rate. The
freedom in choosing the scale $\mu$ reflects our ignorance of
higher-order corrections, which are neglected when the perturbative
expansion is truncated at order $\alpha_s$. Below we shall consider
several choices for the renormalization scale. We allow the pole
masses of the heavy quarks to vary in the range [see (\ref{masses})
and (\ref{mbmcval})]
\begin{equation}
   m_b = (4.8\pm 0.2)~\mbox{GeV} \,, \qquad
   m_b - m_c = (3.40\pm 0.06)~\mbox{GeV} \,,
\end{equation}
corresponding to $0.25<m_c/m_b<0.33$. Non-perturbative effects
appearing at order $1/m_b^2$ in the heavy-quark expansion are
described by the single parameter $\lambda_2\simeq 0.12~\mbox{GeV}^2$
defined in (\ref{mupimuG}); the dependence on the parameter
$\lambda_1$ is the same for all inclusive decay rates and cancels out
in $B_{\rm SL}$ and $n_c$. For the two choices $\mu=m_b$ and
$\mu=m_b/2$, we obtain
\begin{eqnarray}
   B_{\rm SL} &=& \cases{
    12.0\pm 1.0 \% ;& $\mu=m_b$, \cr
    10.9\pm 1.0 \% ;& $\mu=m_b/2$, \cr} \nonumber\\
   \phantom{ \bigg[ }
   n_c &=& \cases{
    1.20\mp 0.06 ;& $\mu=m_b$, \cr
    1.21\mp 0.06 ;& $\mu=m_b/2$. \cr}
\end{eqnarray}
The uncertainties in the two quantities, which result from the
variation of $m_c/m_b$ in the range given above, are anticorrelated.
Notice that the semileptonic branching ratio has a stronger scale
dependence than $n_c$. By choosing a low renormalization scale,
values $B_{\rm SL}<12\%$ can easily be accommodated. The experimental
data prefer a scale $\mu/m_b\sim 0.5$, which is indeed not unnatural.
Using the BLM scale setting method~\cite{BLM}, Luke et al.\ have
estimated that $\mu\gsim 0.32 m_b$ is an appropriate scale in this
case~\cite{LSW}.

\begin{figure}[htb]
   \epsfxsize=8cm
   \centerline{\epsffile{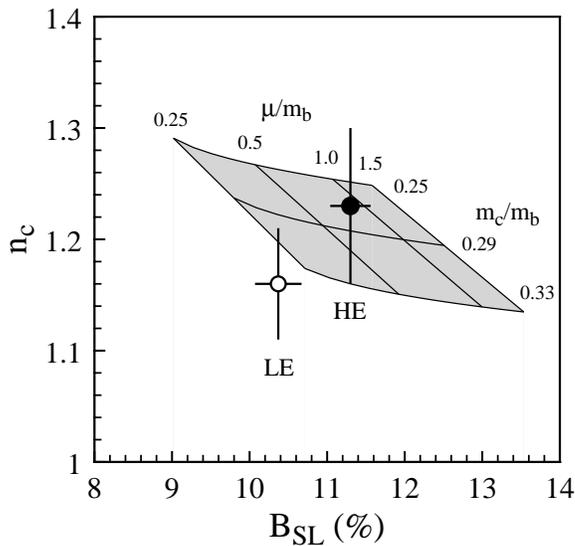}}
\caption{\label{fig:BSL}
Combined theoretical predictions for the semileptonic branching ratio
and charm counting as a function of the quark-mass ratio $m_c/m_b$
and the renormalization scale $\mu$. The data points show the average
experimental values for $B_{\rm SL}$ and $n_c$ obtained in low-energy
(LE) and high-energy (HE) measurements, as discussed in the text.}
\end{figure}

The combined theoretical predictions for the semileptonic branching
ratio and charm counting are shown in Fig.~\ref{fig:BSL}. They are
compared with the experimental results obtained from low- and
high-energy measurements. It was argued that the combination of a low
semileptonic branching ratio and a low value of $n_c$ would
constitute a potential problem for the Standard Model~\cite{Buch}.
However, with the new experimental and theoretical numbers, only for
the low-energy measurements a small discrepancy remains between
theory and experiment. Note that, using (\ref{ncdef}), our results
for $n_c$ can be used to obtain a prediction for the branching ratio
$B(\bar B\to X_{c\bar c s'})$, which is accessible to a direct
experimental determination. Our prediction of $(22\pm 6)\%$ for this
branching ratio agrees well with the preliminary result reported by
the CLEO Collaboration~\cite{Hons}: $B(\bar B\to X_{c\bar c
s'})=(23.9\pm 3.8)\%$.

\section{Concluding Remarks}
 
We have presented an introduction to the theory and phenomenology of
heavy-flavour physics. The theoretical tools that allow us to perform
quantitative calculations in this area are heavy-quark symmetry, the
heavy-quark effective theory, and the $1/m_Q$ expansion. After
presenting the underlying theoretical concepts, we have discussed
their application to the description of exclusive and inclusive weak
decays of $B$ mesons. Besides presenting the status of the latest
developments, our hope was to convince the reader that heavy-flavour
physics is a rich and diverse area of research, which is at present
characterized by a fruitful interplay between theory and experiments.
This has led to many significant discoveries and developments on both
sides. Heavy-quark physics has the potential to determine many
important parameters of the electroweak theory and to test the
Standard Model at low energies. At the same time, it provides an
ideal laboratory to study the nature of non-perturbative phenomena in
QCD, still one of the least understood properties of the Standard
Model.

Let us finish with a somewhat philosophical remark: In the last few
years, there have been exciting developments in high-energy physics
related to the discovery dualities, which relate apparently very
different theories to each other. So are electric, weak-coupling
phenomena in one theory dual to magnetic, strong-coupling phenomena
in another theory. Some people argue quite convincingly that duality
seems to be everywhere in nature, and consequently there are no
really difficult questions in physics; very difficult problems become
trivial when approached from a different, dual point of view. There
are, however, ``moderately difficult'' problems in physics, which are
``self-dual''. It is the author's opinion that real-world (i.e.\
non-supersymmetric) QCD at hadronic energies belongs to this
category. Having said this, we conclude that heavy-quark effective
theory provides a powerful tool to tackle the ``moderately
difficult'' problems of heavy-flavour physics.

\section*{Acknowledgments}
It is my pleasure to thank the organizers of the 20th Johns Hopkins
Workshop for the invitation to present this talk and for making my
stay in Heidelberg such an enjoyable one.

\end{document}